\documentclass[12pt,preprint]{aastex}

\newcommand{\myemail}{handong@mail.tsinghua.edu.cn}

\shorttitle{Comparison between WFFT and HHT} \shortauthors{Han \&
Zhang}

\begin{document}

\title{Comparison between Windowed FFT and
Hilbert-Huang Transform for Analyzing Time Series with Poissonian
Fluctuations: A Case Study }

\author{Dong Han\altaffilmark{1} and Shuang-Nan Zhang\altaffilmark{2,~ 3}}
\altaffiltext{1}{Department of Engineering Physics \& Center for
Astrophysics, Tsinghua University, Beijing 100084; \myemail}
\altaffiltext{2}{Department of Physics \& Center for Astrophysics,
Tsinghua University, Beijing 100084;
zhangsn@mail.tsinghua.edu.cn}
\altaffiltext{3}{Key Laboratory of Particle Astrophysics,
Institute of High Energy Physics, Chinese Academy of Sciences,
Beijing 100049}

\begin{abstract}
Hilbert-Huang Transform (HHT) is a novel data analysis technique
for nonlinear and non-stationary data. We present a time-frequency
analysis of both simulated light curves and an X-ray burst from
the X-ray burster 4U 1702-429 with both the HHT and the Windowed
Fast Fourier Transform (WFFT) methods. Our results show that the
HHT method has failed in all cases for light curves with
Poissonian fluctuations which are typical for all photon counting
instruments used in astronomy, whereas the WFFT method can
sensitively detect the periodic signals in the presence of
Poissonian fluctuations; the only drawback of the WFFT method is
that it cannot detect sharp frequency variations accurately.

\end{abstract}


\section{INTRODUCTION}

In X-ray binaries, matter flows onto the neutron star or black
hole through an accretion disk and X-rays are emitted by the
accretion flow from inner regions that host the strongest
gravitational fields in the Universe. The space time
characteristics of the X-ray emitting regions lead to the
conclusion that the dynamical time scale for the accretion process
of X-ray emitting is millisecond. That means, the millisecond
X-ray variability is caused by orbital motion, neutron-star spin,
disk- and neutron-star oscillations. So the millisecond X-ray
variabilities are important for studying strong gravity physics.
Kilohertz quasi-periodic oscillations (kHz QPOs) , millisecond
pulsations and burst oscillations are millisecond X-ray
variabilities. Burst oscillations do not occur in every X-ray
burst and are different from burst to burst \citep{klis00}.

Fourier transformation is a traditional method in the
time-frequency analysis of linear, stationary and global time
series. In astrophysics, Fourier spectral analysis is also widely
used in the time-frequency analysis, although time series in
astronomy are sometimes non-stationary. Based on the traditional
Fourier spectral analysis, Windowed-Fast-Fourier Transform (WFFT)
assumes that the time series to be piecewise stationary. In order
to result in good time resolution, the window width of WFFT must
be narrow; on the other hand, to achieve a high frequency
resolution and high detection sensitivity, a wide window width is
required. Therefore, in practice the window width must be chosen
carefully depending on the goal of the study.

Both kHz QPOs and burst oscillations were detected in RXTE
observations of the atoll source 4U 1702-429 in 1997. There were
six type I X-ray burst, with a rise time of $<$ 1 s, and a 10-100
s exponential decay. Time-frequency distributions of the light
curve from 4U 1702-429 at time resolution 1/8192 s were presented
with WFFT with the unit signal $u(t)$ as the window function, and
a window width of 2 s. All power spectra were computed by
successively sliding the window at a step of 0.125 s by
\citet{mark99}. They discovered burst oscillations at about 330 Hz
in five of the six type I X-ray bursts. In the 1997 July 26 burst
a periodic signal was observed whose frequency increased slowly
from 328 Hz to 330 Hz.

The advantages of the Hilbert-Huang Transform (HHT)
\citep[see][]{huang98} are that it can be used to analyze
nonlinear and non-stationary data without specifying a basis and
the data decomposition is made only on physical grounds, compared
to Fourier decomposition and wavelet decomposition. Because the
light curves obtained with most astronomical instruments involving
photon counting devices are characterized by Poissonian
fluctuations, it will be very useful if we can analyze such light
curves with HHT and obtain detailed physical information such as
fast frequency variations. Now, for the light curve of 4U
1702-429, the calculated time-frequency distribution and
energy-frequency-time distribution using HHT are quite different
from those using WFFT. In this paper we shall investigate the
large discrepancies from these two methods by making various
numerical tests. The structure of the paper is as follows. In
section 2, we introduce the concepts and definitions of {\it
Fourier frequency} and {\it instantaneous frequency} in the
Hilbert Transform. In section 3, we introduce the HHT. In section
4, we make comprehensive comparisons between the HHT and WFFT.
Finally we summarize our results and present a discussion in
section 5.

\section{FOURIER FREQUENCY AND INSTANTANEOUS FREQUENCY}

Fourier Transform is very important in time series analysis. The
bases of Fourier expansion are trigonometric functions. The
resulted frequency from Fourier decomposition is called {\it
Fourier frequency}.

For an arbitrary time series, $x(t)$, its Hilbert transform $y(t)$
can be expressed as
\begin{equation}
y(t)= \frac{1} { \pi} \int \frac{x( \tau)} {(t- \tau)} d \tau,
\end{equation}
so an analytic signal, $z(t)$, is defined in terms of $x(t)$ and
$y(t)$ as
\begin{equation}
z(t)= x(t)+iy(t)=a(t)e^{i \theta(t)},
\end{equation}
where
\begin{displaymath}
a(t) = \sqrt{ x^2(t)+y^2(t)}
\end{displaymath}
\begin{displaymath}
\theta(t) = \arctan{ \frac{y(t)} {x(t)}}.
\end{displaymath}
The {\it instantaneous frequency} is defined as
\begin{equation}
f(t)=  \frac{1} {2 \pi} \theta \arcmin (t).
\end{equation}
which is the result of the analytic function by Hilbert transform.
It is the first derivative of phase and thus different from the
Fourier frequency. The Fourier frequency is constant over the
whole time series, while the instantaneous frequency is a function
of time. In other words, since all the trigonometric bases of the
Fourier transform contribute to Fourier spectrum globally, the
Fourier frequency is the global result of the time series within a
time span, whereas the instantaneous frequency is a local property
of the time series. Thus, only when the time series is composed of
pure trigonometric functions, are the Fourier frequencies equal to
the instantaneous frequencies, and the instantaneous frequencies
of the trigonometric functions constant over the whole time series
\citep{hu04}.

\section{THE HILBERT HUANG TRANSFORM}

The Hilbert Huang Transform was developed by
\citet{huang98,huang99}. In this method any time series, including
non-linear and non-stationary series, can be decomposed into a
finite number of \emph{intrinsic mode functions} (IMFs) through
\emph{empirical mode decomposition} (EMD) process. An IMF is a
function which must satisfy the following two conditions: (1) the
difference between the numbers of extrema and zero-crossings is
$\leq$ 1; and (2) the mean of the upper envelop (linked by local
maxima) and the lower envelop (linked by local minima) is zero at
every point. With Hilbert transform, the IMFs yield
energy-frequency-time distribution, the so-called Hilbert
spectrum. HHT has been studied extensively, and has been widely
applied in many fields of research.
\citep{nair03,zim02,vara04,huang03b}

The EMD process is as follows. According to HHT, once the extrema
of a time series $x(t)$ are identified, all the local maxima and
minima are connected by two special lines, called the upper and
lower envelopes. Their mean is designated as $m_1$, and the
difference between $x(t)$ and $m_1$ is $x(t)-m_1=h_1$. If $h_1$ is
not an IMF, $h_1$ is treated as the data and undergos the
procedure above, giving $h_1-m_{11}=h_{11}$. Repeat this sifting
procedure $k$ times until $h_{1k}$ is an IMF, that is
$h_{1(k-1)}-m_{1k}=h_{1k}$, thus the first IMF component is
obtained, i.e. IMF$_1=h_{1k}$. Then separate IMF$_1$ from the
original time series by $x(t)-$IMF$_1=r_1$. Treat $r_1$ as the new
data and subject it to the same sifting process above. Repeat this
procedure on all the subsequent $r_j$s, i.e.
$r_1-$IMF$_2=r_2,...,r_{n-1}-$IMF$_n=r_n$. So the result is
\begin{equation}
x(t)= \sum^n_{j=1}{IMF_j(t)+r_n(t)}.
\end{equation}
After Hilbert transform on each IMF, the data can be expressed as
\begin{equation}
x(t)=\sum^n_{j=1}{a_j(t)exp(i \theta_j(t))},
\end{equation}
then, the instantaneous frequency is
\begin{equation}
\omega_j(t)=  \frac{1} {2 \pi} \frac {d \theta_j(t)} {dt}.
\end{equation}
So the final result is the distribution of the amplitude (of
energy) and frequency as the functions of time.

If the time length of $x(t)$ is $T$ and the sampling rate is
$\triangle t$, the time resolution is $\triangle t$ and the lowest
extractable frequency is $1/T$, which is also the best frequency
resolution. The frequency resolution is up to the Nyquist
frequency.

\section{NUMERICAL COMPARISON BETWEEN WFFT AND   HHT}

Different time series, which will be described in the following
subsections, are analyzed with HHT-DPS 1.3b, which is a free
test-version software from the open NASA internet web site. The
digitizing rate of the time series is 1/8192 s and the total time
length is 10 s, the same as that of the light curve of the X-ray
burst from 4U 1702-429. So the time resolution of the result with
HHT is 1/8192 s and the frequency resolution is 0.1 Hz, while the
time resolution of WFFT is 0.125 s and the frequency resolution is
0.5 Hz.

If we set the time and frequency resolutionS of HHT the same as
those of WFFT, the time-frequency-energy distribution will be too
illegible to reveal anything. The resolution of the computed
energy-frequency-time distribution with HHT-DPS 1.3b is $400
\times 400$ for the time and frequency axis, i.e. the time span
from 0-10 s is divided into 400 parts, and so is the computed
frequency range from 0 to the maximal frequency. The time
resolution of the computed time-frequency distribution with
HHT-DPS 1.3b is 1/8192 s, so the illustrated time-frequency
distributions with HHT-DPS 1.3b are smoothed with a 205-point
average to meet the time resolution of 400 divisions.

\subsection{Comparison between HHT and WFFT with Simple Input Periodic Functions}
\label{bozomath}

We compare between WFFT and HHT using four examples of periodic
functions with amplitude set at 3000, approximately the same as
the X-ray burst from 4U 1702-429. The input time series and the
calculated results with HHT and WFFT are listed in
Table~\ref{tbl-1}.

\begin{deluxetable}{ccrrrrrrrrcrl}
\tablecaption{Testing HHT and WFFT with Simple Input Periodic
Functions\label{tbl-1}} \tablewidth{0pt} \tablehead{
\colhead{Input functions} & \colhead{Results with HHT} &
\colhead{Results with WFFT}} \startdata $ y=3000\times(\sin(700
\pi
t)+\sin(424 \pi t)$ & & \\
  $+\sin(300 \pi t)+3)$. & Fig. 1 & Left of Fig. 5\\
Function as above, but with Poissonian sampling. &
Fig. 2 & Right of Fig. 5\\
$y=\left\{%
\begin{array}{ll}
    3000 \times (\sin(300 \pi t)+1), & \hbox{t $<$ 5 s} \\
    3000 \times (\sin(700 \pi t)+1), & \hbox{t $\geq$ 5 s}. \\
\end{array}%
\right. $ & Fig. 3 & Left of Fig. 6\\
Function as above, but with Poissonian sampling. & Fig. 4 & Right
of Fig. 6

\enddata

\end{deluxetable}

Figure 1 shows the calculated results with HHT for a time series
composed of three sinusoidal functions and a low-level
DC-component. Clearly the HHT method failed to detect the three
frequencies correctly. For the above times series sampled with
Poissonian fluctuations, the HHT method can detect all three
sinusoidal components, as shown in Figure 2. However, large spikes
in the detected frequencies exist, thus making the {\it
instantaneous frequency} meaningless. In comparison as shown in
Figure 5, the WFFT can detect all three sinusoidal components
correctly with or without Poissonian fluctuations as expected.

In Figures 3 and 4, we show the calculated results with HHT for a
time series with just one sinusoidal component whose frequency has
a sudden jump in the middle of the time series. The HHT method can
detect the sinusoidal component correctly; in the presence of
Poissonian fluctuations large spikes are again found, however,
which make it difficult to judge if the spikes are intrinsic in
the time series or simply due to fluctuations. In comparison, as
shown in Figure 6, the WFFT method can detect the component
correctly without spurious spikes; but the time accuracy of the
frequency jump is reduced to about 0.5 s because of the window
applied in doing the FFT.

\subsection{A Case Study of an X-ray Burst from 4U 1702-429 with HHT and WFFT} \label{bozomath}

We test HHT with the light curve of an X-ray burst from 4U
1702-429 in 1997, which is shown as the solid line in the left
panel of Fig. 7. The power spectra (shown as contours) of the
original light curve produced with WFFT are also illustrated in
the left panel of Figure 7; clearly a periodic component with a
varying frequency between 328 Hz to around 330 Hz is detected,
confirming the results of \citet{mark99}. In comparison, the
calculated time-frequency distribution and energy-frequency-time
distribution with HHT are shown in Fig. 8. The computed
time-frequency distribution with HHT is composed of a series of
frequency components with energy generally increasing from low
frequency components to high frequency components. The highest
frequency about 3000 Hz, due to Poissonian fluctuations, is
physically meaningless, but has the largest energy intensity. The
presented results with HHT are completely different from those
with WFFT from Figure 7 (left panel): (1) There is not an obvious
high intensity frequency component at about 330 Hz, as detected
with WFFT; (2) The movement of frequencies with time is also
different from that with WFFT; and (3) the fluctuations in
frequency components make it impossible to reveal any real
periodic signals.

\subsubsection{ Test HHT-DPS with synthesized light curves}\label{bozomath}

To understand why the above calculated results with HHT and WFFT
are so different, we take the following steps to find out the
conditions at which WFFT or HHT can detect real periodic signals
in time series with Poissonian fluctuations.

We eliminate the periodic components in the observed light curve
of the X-ray burst, by a smoothening treatment for the light curve
through a 25-point average. Then, the smoothed light curve is
processed with Poissonian sampling, and the new light curve is
analyzed with HHT-DPS 1.3b, in order to understand the behaviors
of HHT for Poissonian time series with varying intensities but
without periodic signals. To study the minimum strength of
periodic signals HHT can detect reliably for the above light
curve, a sinusoidal signal is added to the smoothed light curve,
and the amplitude and frequency of the added-in sinusoidal signal
are both kept constant during the whole time span. Calculations
for the synthesized light curves with amplitudes of 0.1 and 0.4
times the mean value of the light curve with a frequency of 400 Hz
are made with HHT. All the results obtained with HHT are compared
with those with WFFT. All input synthesized light curves and the
calculated results are listed in Table~\ref{tbl-2}.

\begin{deluxetable}{lrrrrcrrrrr}
\tablecaption{Comparison between HHT and WFFT with Synthesized
Light Curves\label{tbl-2}} \tablewidth{0pt} \tablehead{
\colhead{Synthesized light curve} & \colhead{Results with HHT} &
\colhead{Results with WFFT}} \startdata
The 25-point smoothed light curve & & \\
with Poissonian sampling. &
Fig. 9 & Right of Fig. 7\\
The smoothed light curve with added-in & & \\
signal and Poissonian sampling.\tablenotemark{a} &
Fig. 10 & Left of Fig. 12\\
Light curve as above but with & & \\
a different amplitude.\tablenotemark{b} &
Fig. 11 & Right of Fig. 12\\

\enddata

\tablenotetext{a}{Start with a sinusoidal signal with amplitude
0.1 times the mean value of the light curve, and with frequency
set at 400 Hz, add this signal to the smoothed light curve, then
sample with Poissonian sampling.} \tablenotetext{b}{The same
synthesized light curve as above but with a different amplitude of
0.4 times the mean original light curve.}

\end{deluxetable}

The calculated results of WFFT agree with our expectations: (1)
From the right panel of Figure 7, the original frequency component
about 330 Hz disappeared from the smoothed light curve, and (2)
the constant frequency of 400 Hz can be detected clearly for the
two test cases, as shown in Figure 12.

For the results of HHT, comparing Figure 9 with Figure 8, we can
see, as expected, that the energy of the frequency components
higher than 500 Hz becomes weaker when the input time series is
the smoothed light curve. Figures 10 and 11 show that all
frequency components changed when the periodic signal is added in.
However, the 400 Hz frequency component is not recognizable at all
in Figure 10, and barely seen in Figure 11, but the large number
and high energy spikes make it impossible to draw any reliable
conclusion on the frequency component from the HHT results.

\section{CONCLUSIONS}

We have reached the following conclusions: (1) Poissonian
fluctuations cause large spurious spikes in the calculated Hilbert
spectra with the HHT method, sO that in all cases the HHT cannot
detect periodic signals reliably as long as the time series is
essentially Poissonian; and (2) in comparison, the WFFT method is
relatively insensitive to Poissonian fluctuations and can detect
periodic signals in all cases with high sensitivity, but the time
of frequency jump, should it exists, calculated with WFFT is less
accurate than that with HHT .

\citet{wu04} established the statistical characteristics of white
noise through empirical study on Fourier spectra of the IMFs of
the numerical white-noise time series with EMD, and provided a
method for finding out which IMF components represent signals and
which represent noises. However, as we have shown in this paper it
is practically impossible to eliminate the spurious spikes in the
calculated Hilbert spectra with HHT for Poissonian time series,
consequently any empirical estimate of the statistical
significance on the frequency component does not help in detecting
underlying periodic signals.

It is worth noting, however, that it is possible to combine the
HHT and WFFT together to extract as much information as possible
from observed time series. For example, the WFFT method may be
used to identify all frequency components in the time series and
study any slow frequency variations. After the WFFT method
identified a frequency jump, as shown in Figure 6, the HHT method
can be used to identify the exact time of the sudden jump, as
shown in Figure 3 and 4.

\acknowledgments We thank the anonymous referee for our previous
publication \citep{feng04}, whose comments on the HHT method
motivated the work reported here. We also thank T. P. Li, M. Wu
and L. M. Song for many interesting discussions. This study is
supported partly by the Special Funds for Major State Basic
Research Projects, the Directional Research Grant of the Chinese
Academy of Sciences, the National Natural Science Foundation under
grant Nos. 10521001 and 10233010, and the Ministry of Education of
China.

\clearpage

\begin{figure}
\plottwo{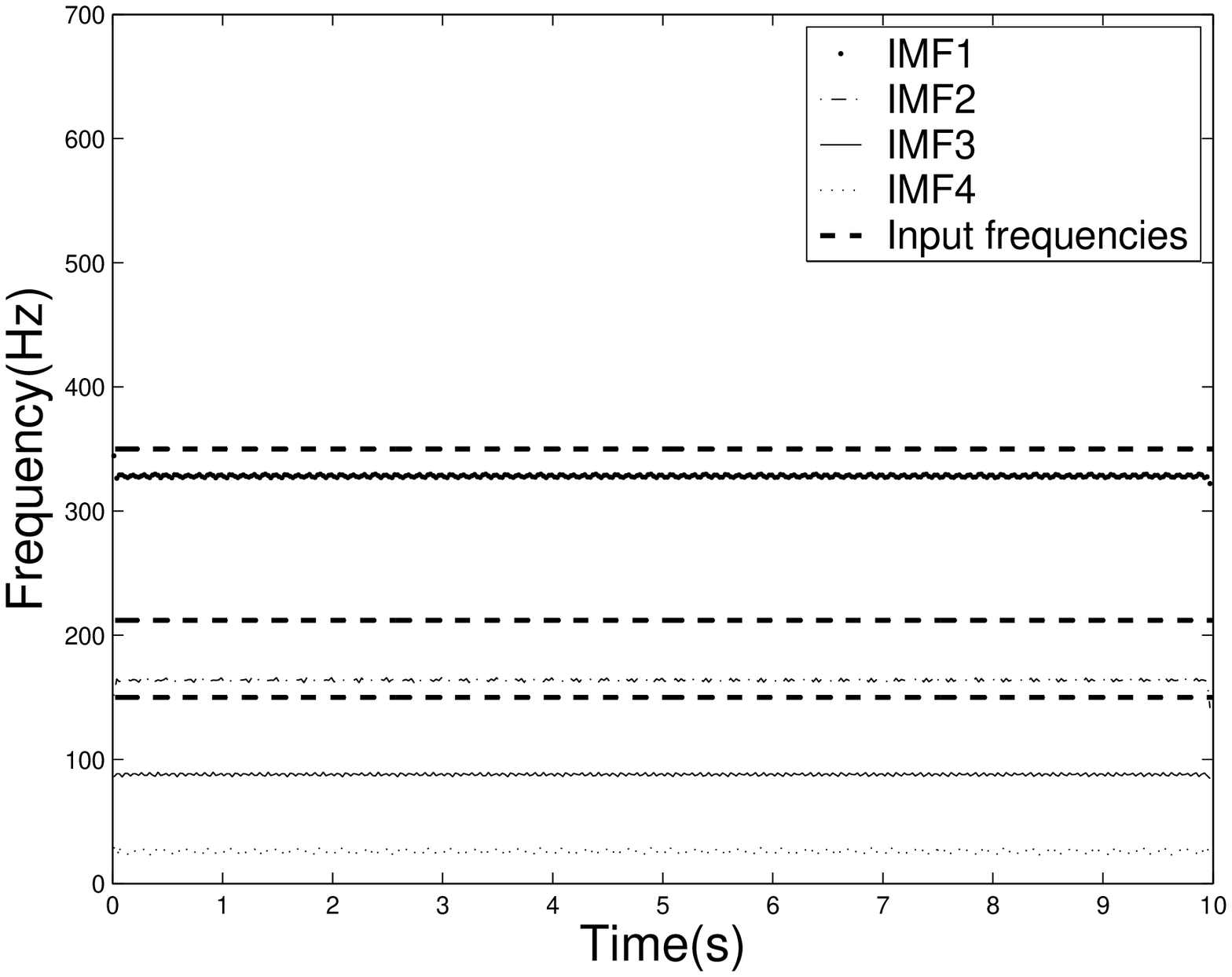}{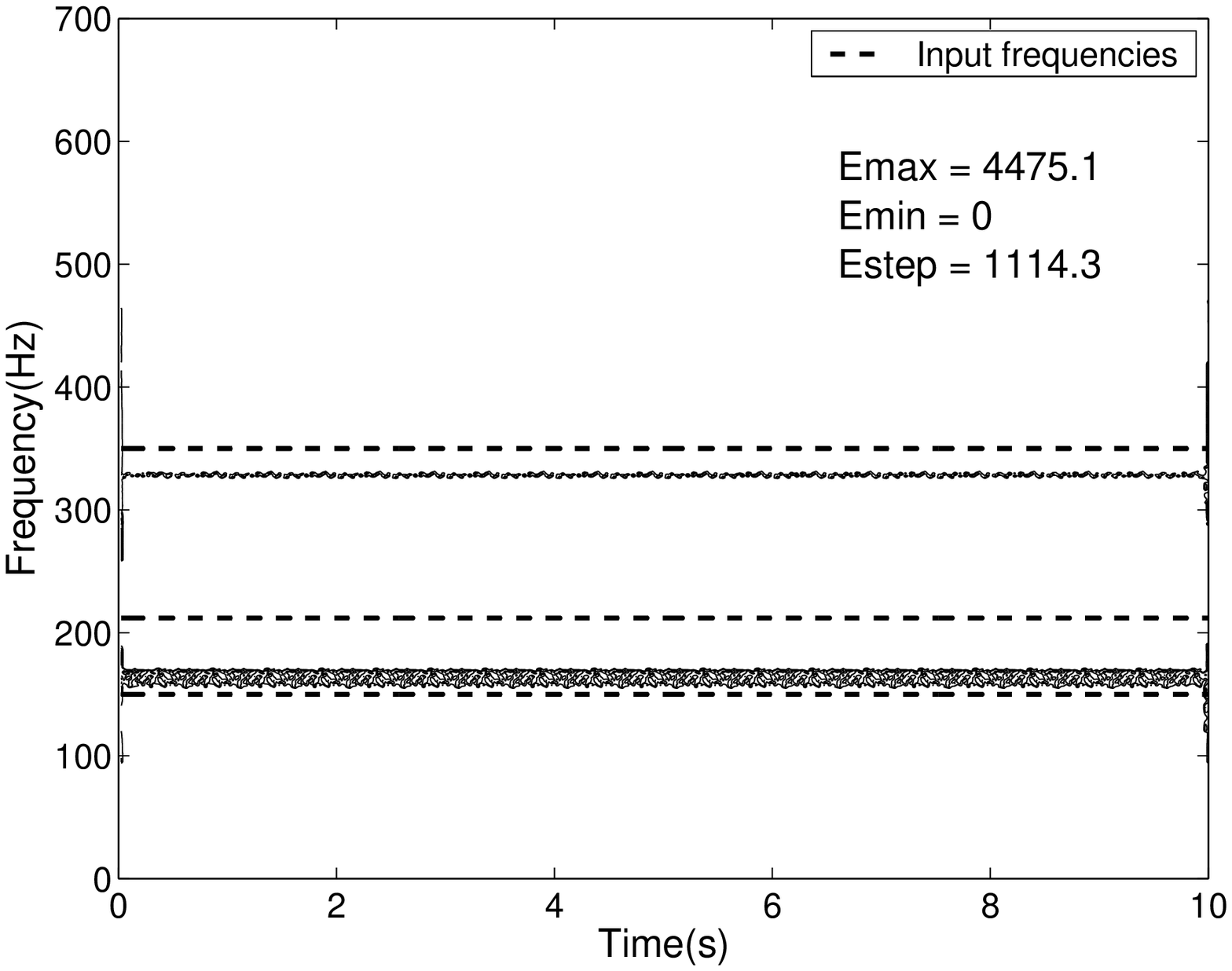} \caption{Calculated time-frequency
distribution (left) and energy-frequency-time distribution (right)
with HHT. The input time series is the sinusoidal signal as listed
in the first row of Table 1. The dashed lines are for 350 Hz, 212
Hz and 150 Hz, which are the frequencies of the input periodic
signals in the time series. For the right panel, the values of the
maximum, minimum and step size of the contours are indicated (this
is also done in all other contour plots).\label{fig-1}}
\end{figure}

\begin{figure}
\plottwo{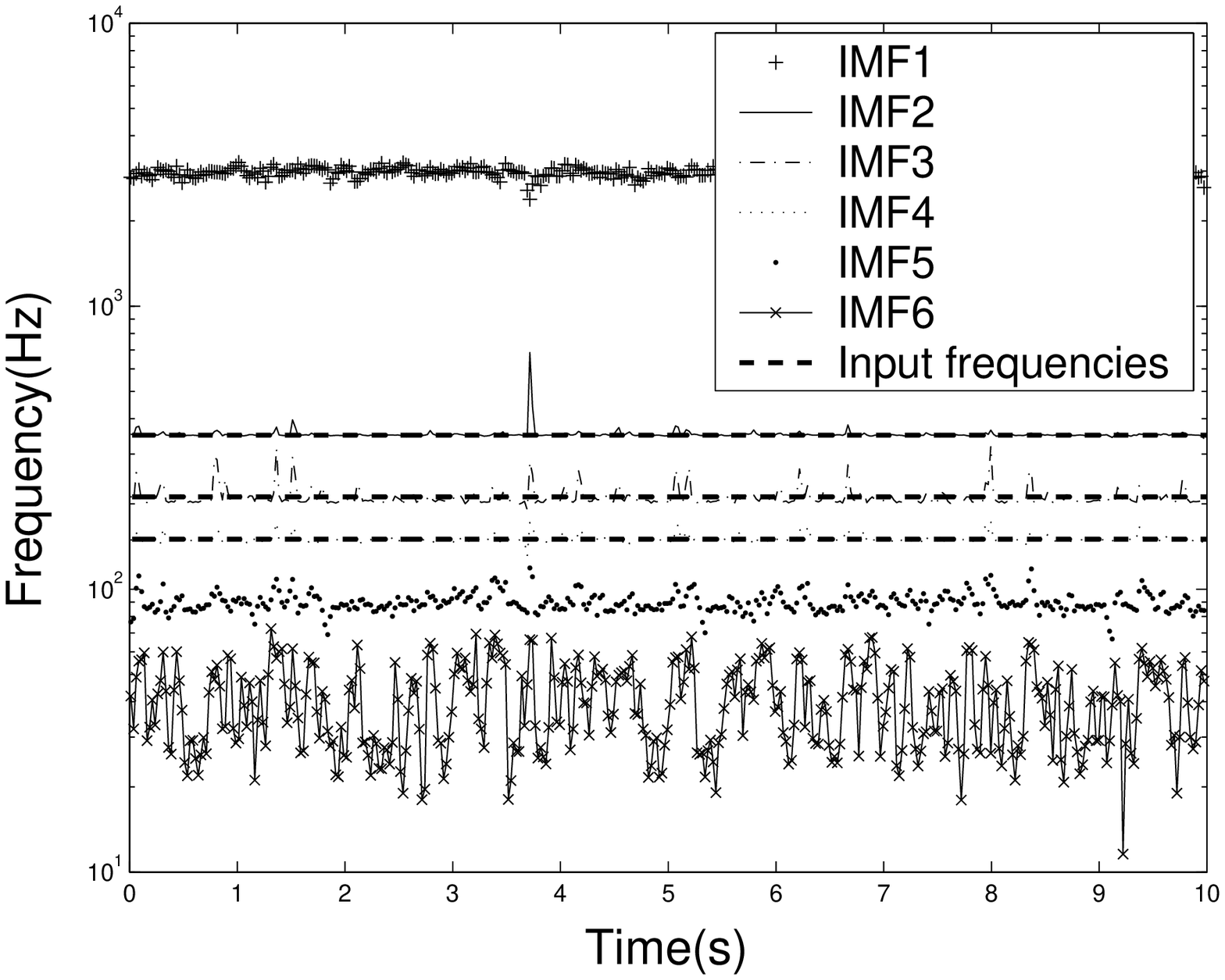}{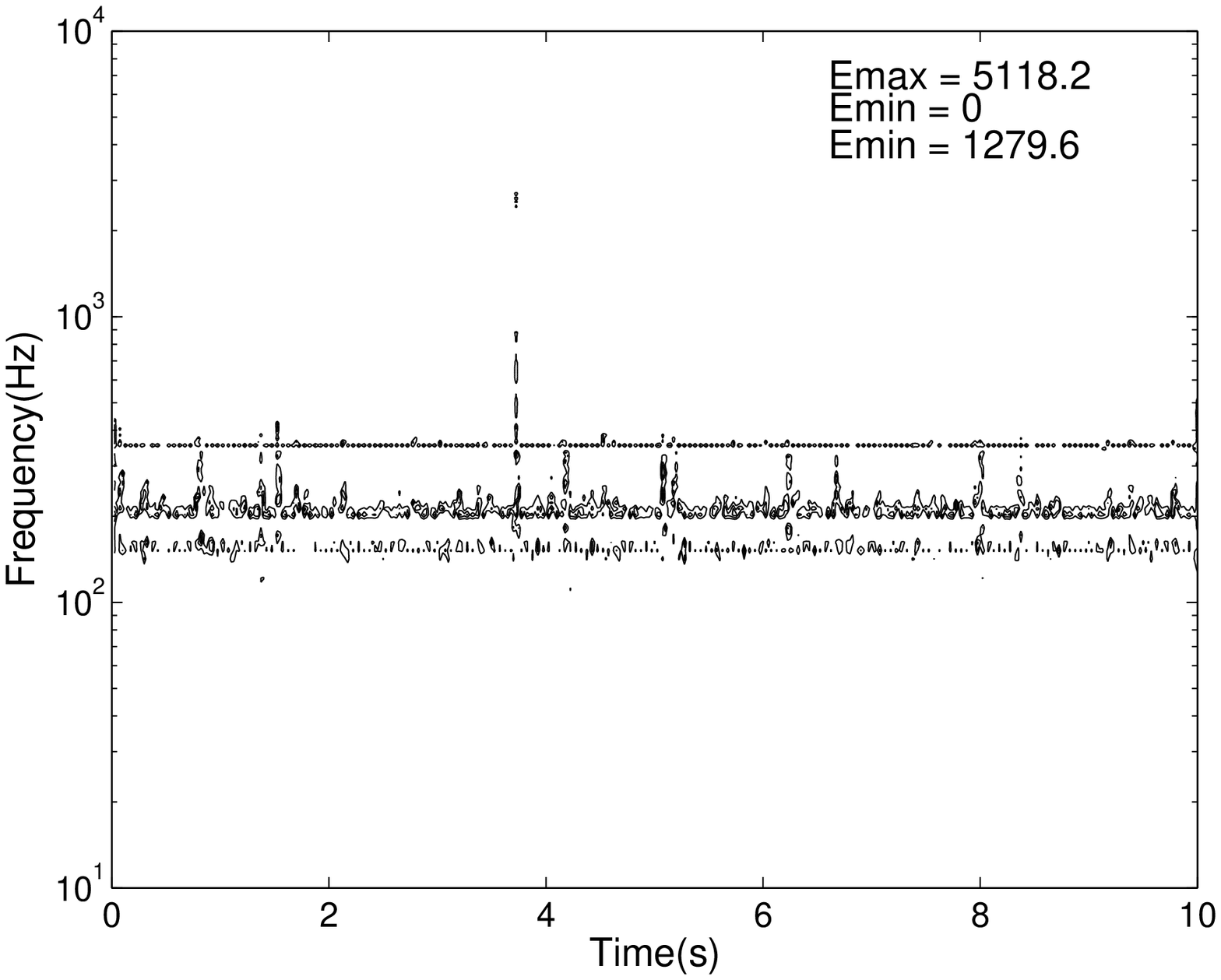} \caption{Calculated time-frequency
distribution (left) and energy-frequency-time distribution (right)
with HHT. The input time series is the same as that of Figure 1,
but with Poissonian sampling as listed in the second row of Table
1. The dashed lines of the left panel are for 350 Hz, 212 Hz and
150 Hz, which are the frequencies of the input periodic signals in
the time series.\label{fig-2}}
\end{figure}

\clearpage

\begin{figure}
\plottwo{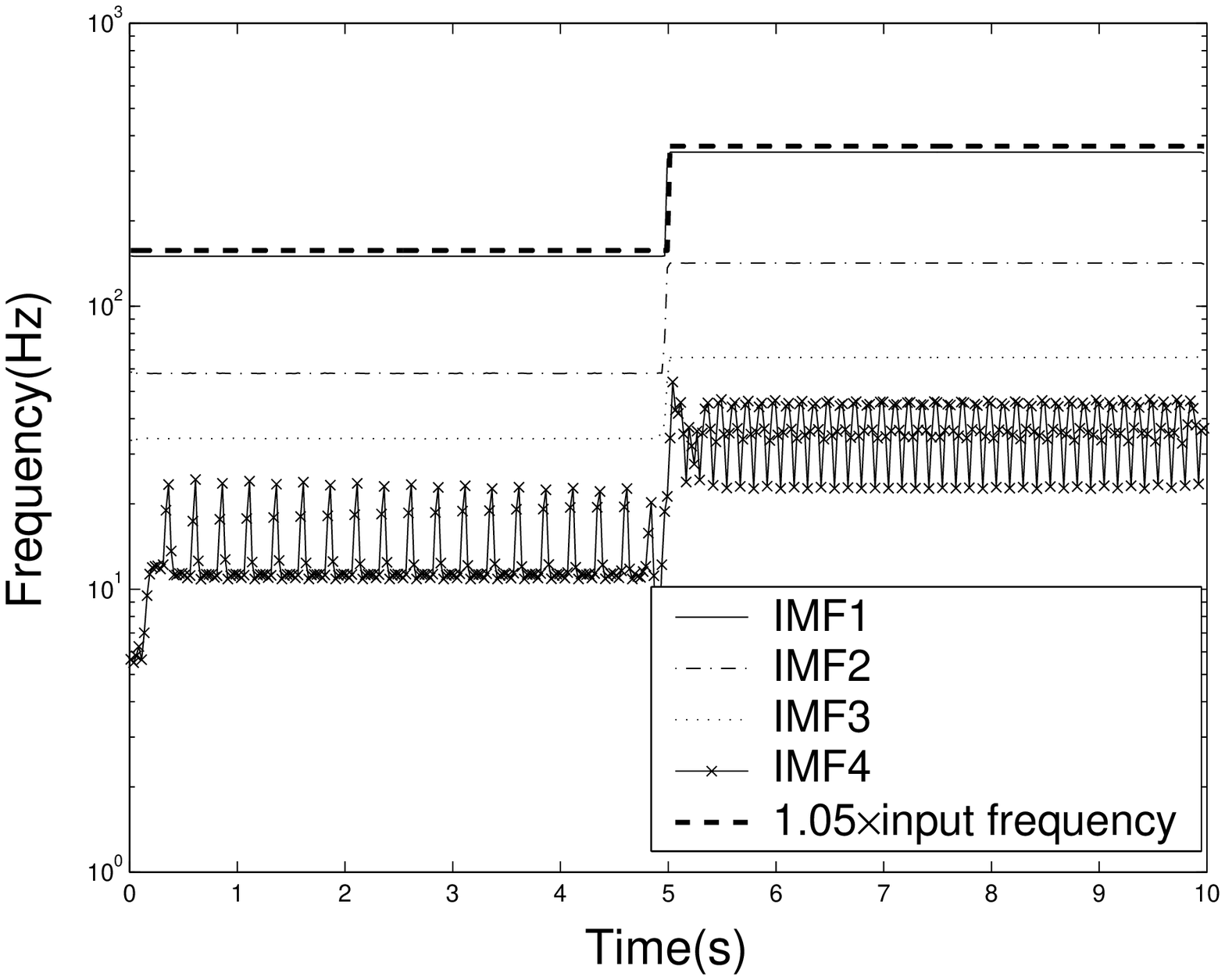}{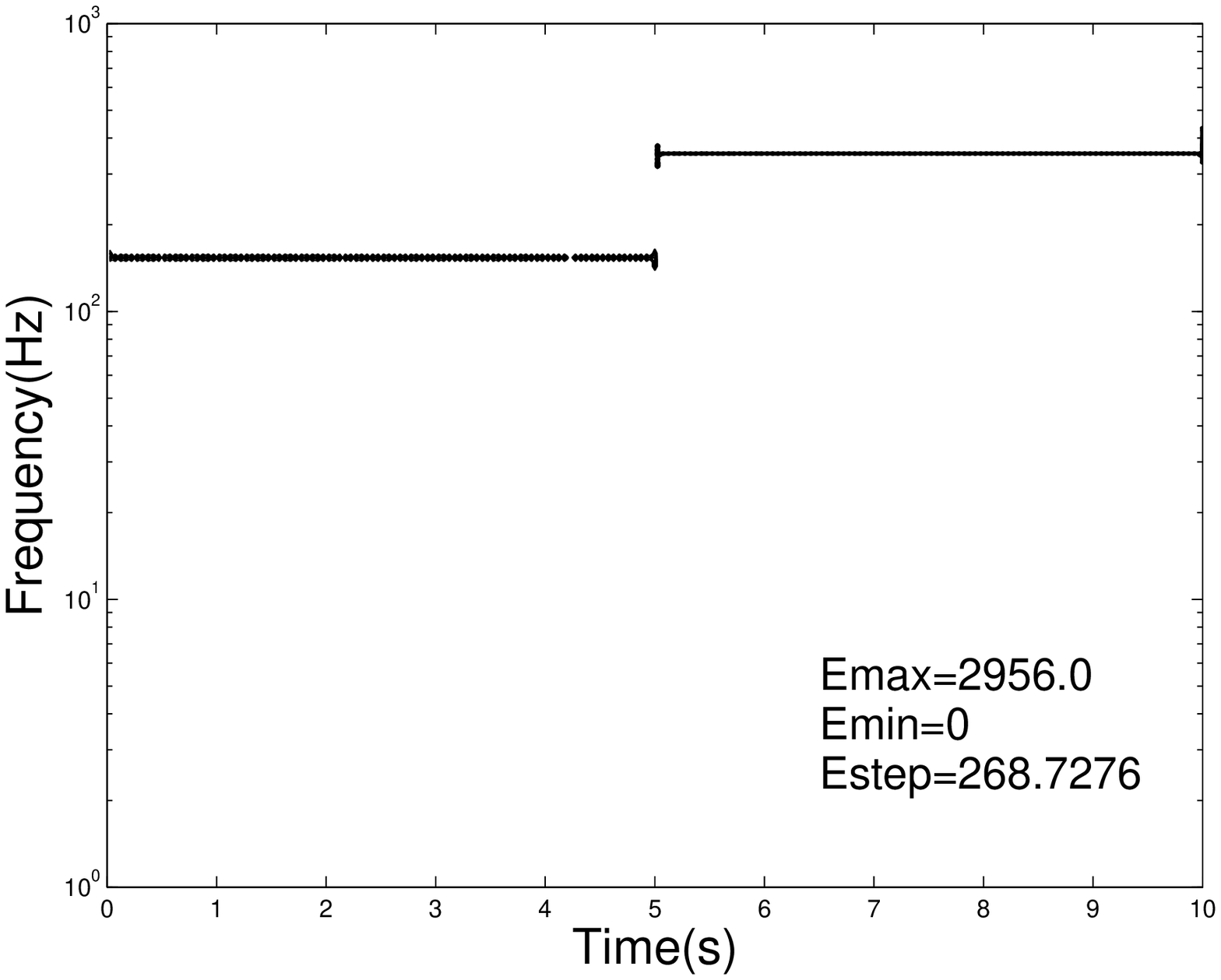} \caption{Calculated time-frequency
distribution (left) and energy-frequency-time distribution (right)
with HHT. The input time series is a period signal with just one
sinusoidal component whose frequencies have a sudden jump at 5 s
as listed in the third row of Table 1. The dashed line of the left
panel is for $1.05\times$ the frequency of the input periodic
signal.\label{fig-3}}
\end{figure}

\begin{figure}
\plottwo{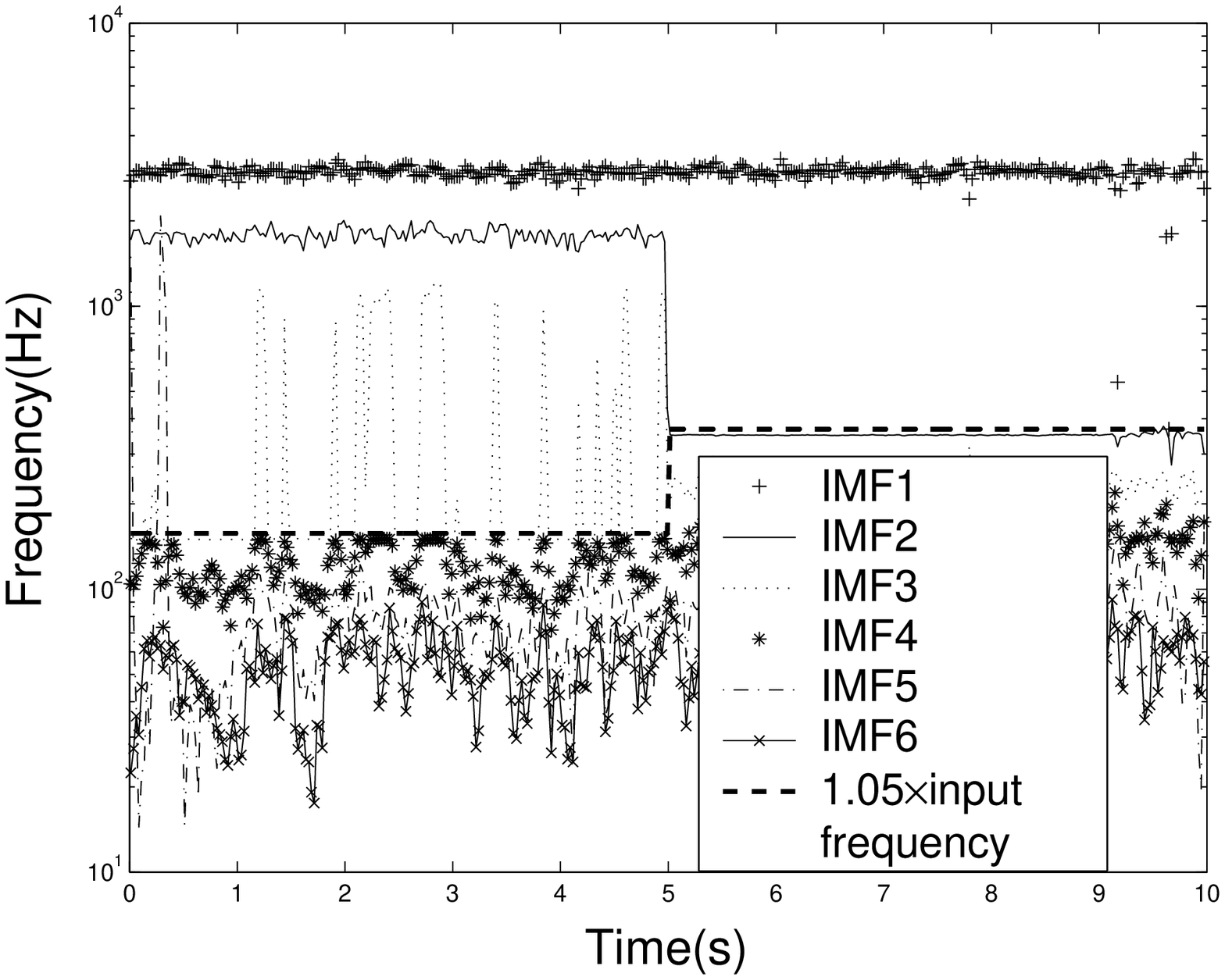}{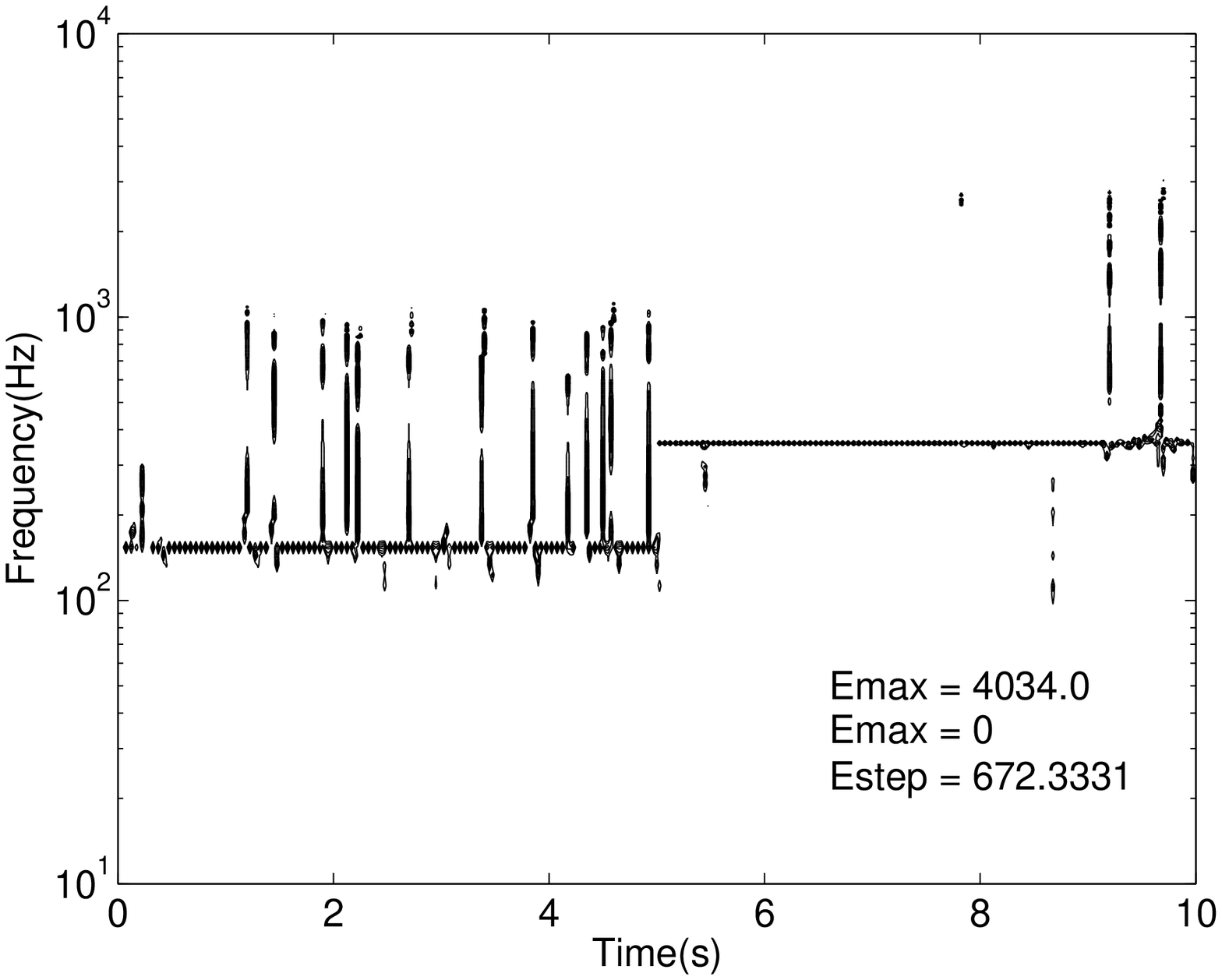} \caption{Calculated time-frequency
distribution (left) and energy-frequency-time distribution (right)
with HHT. The input time series is the same as that of Figure 3,
but with Poissonian sampling as listed in the forth row of Table
1. The dashed line of the left panel is for $1.05\times$ the
frequency of the input periodic signal.\label{fig-4}}
\end{figure}

\clearpage

\begin{figure}
\plottwo{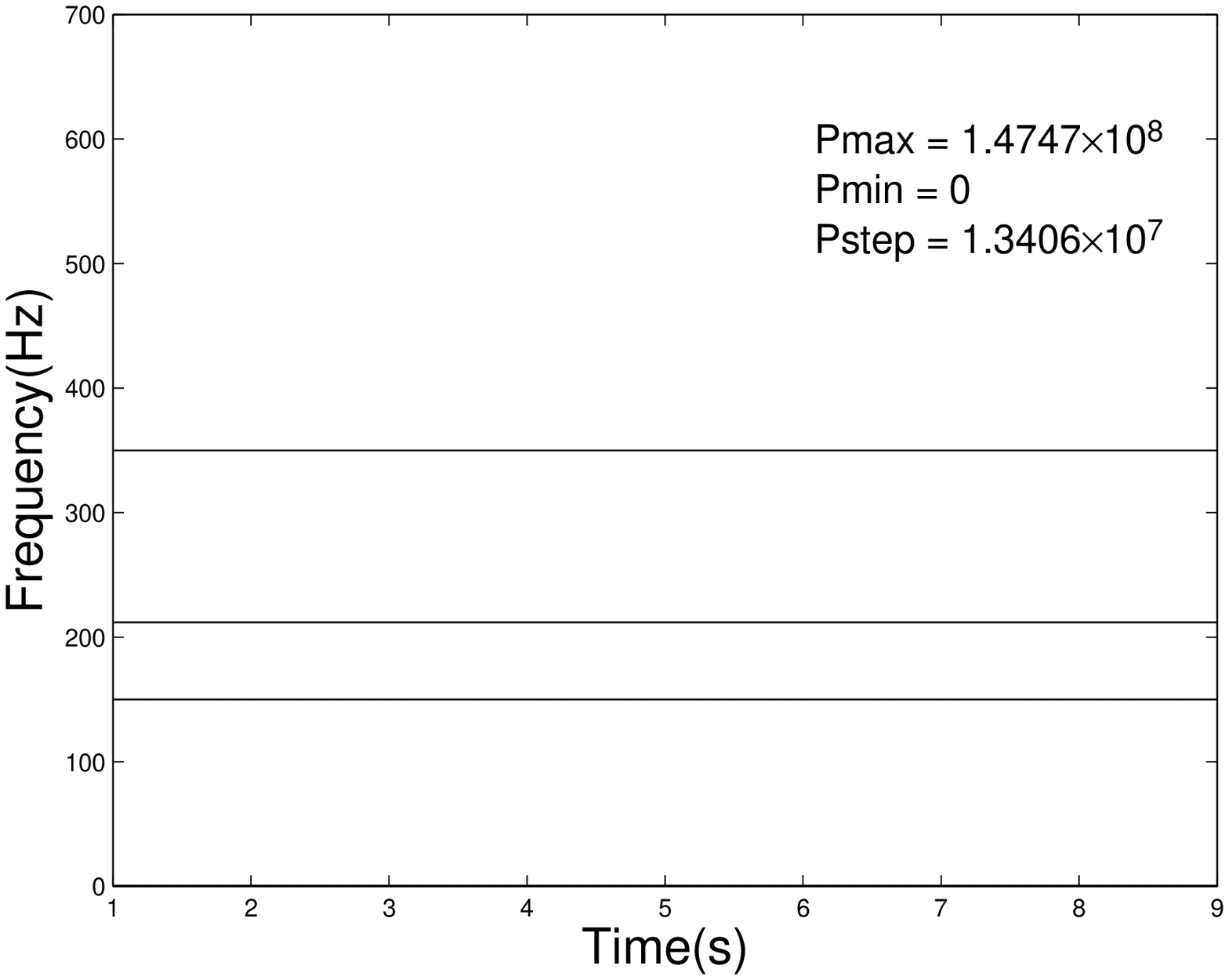}{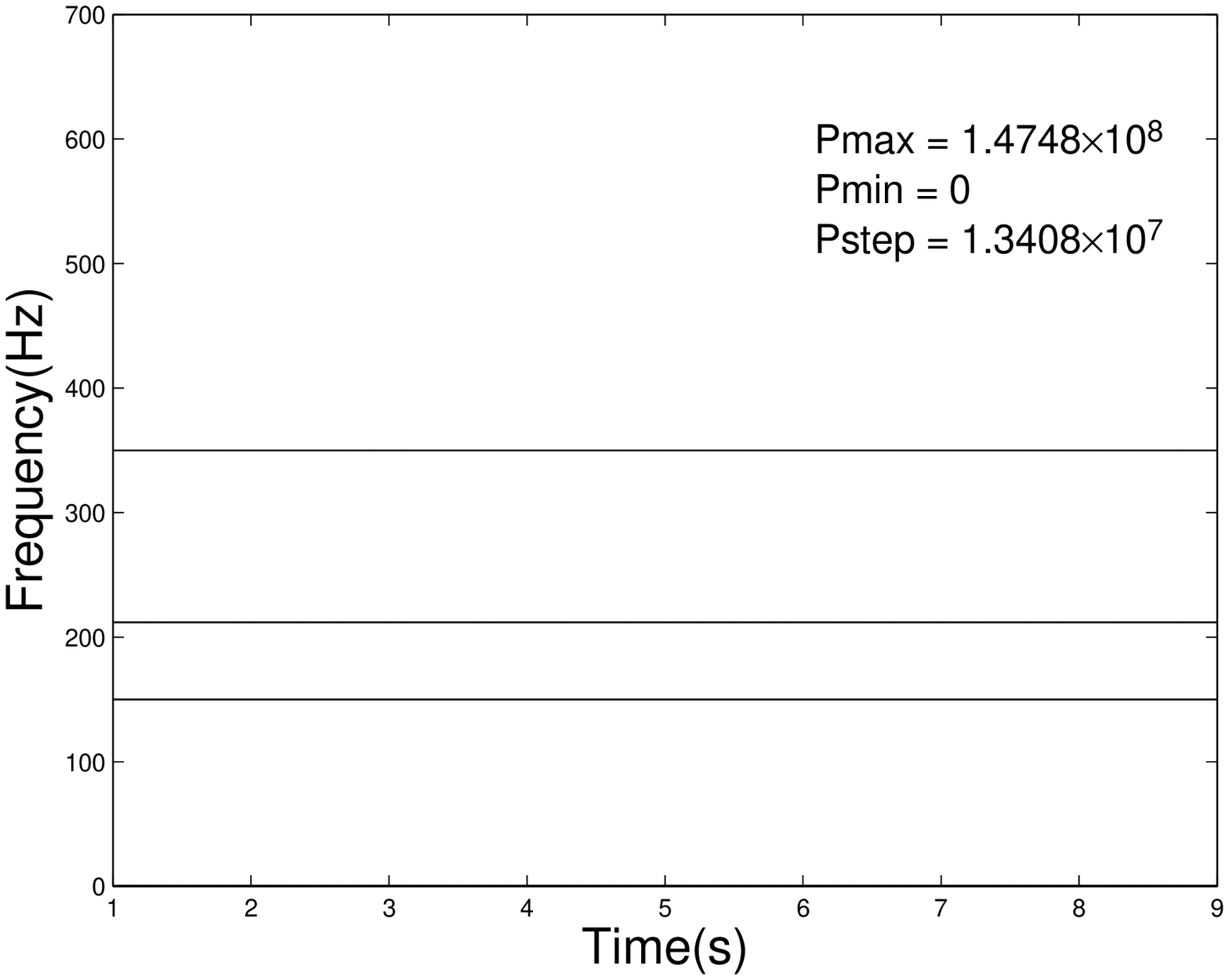} \caption{Calculated power spectra with
WFFT. Left: the result of the input sinusoidal signal as listed in
the first row of Table 1. Right: the result of the same time
series as that of left panel, but with Poissonian sampling as
listed in the second row of Table 1.\label{fig-5}}
\end{figure}

\begin{figure}
\plottwo{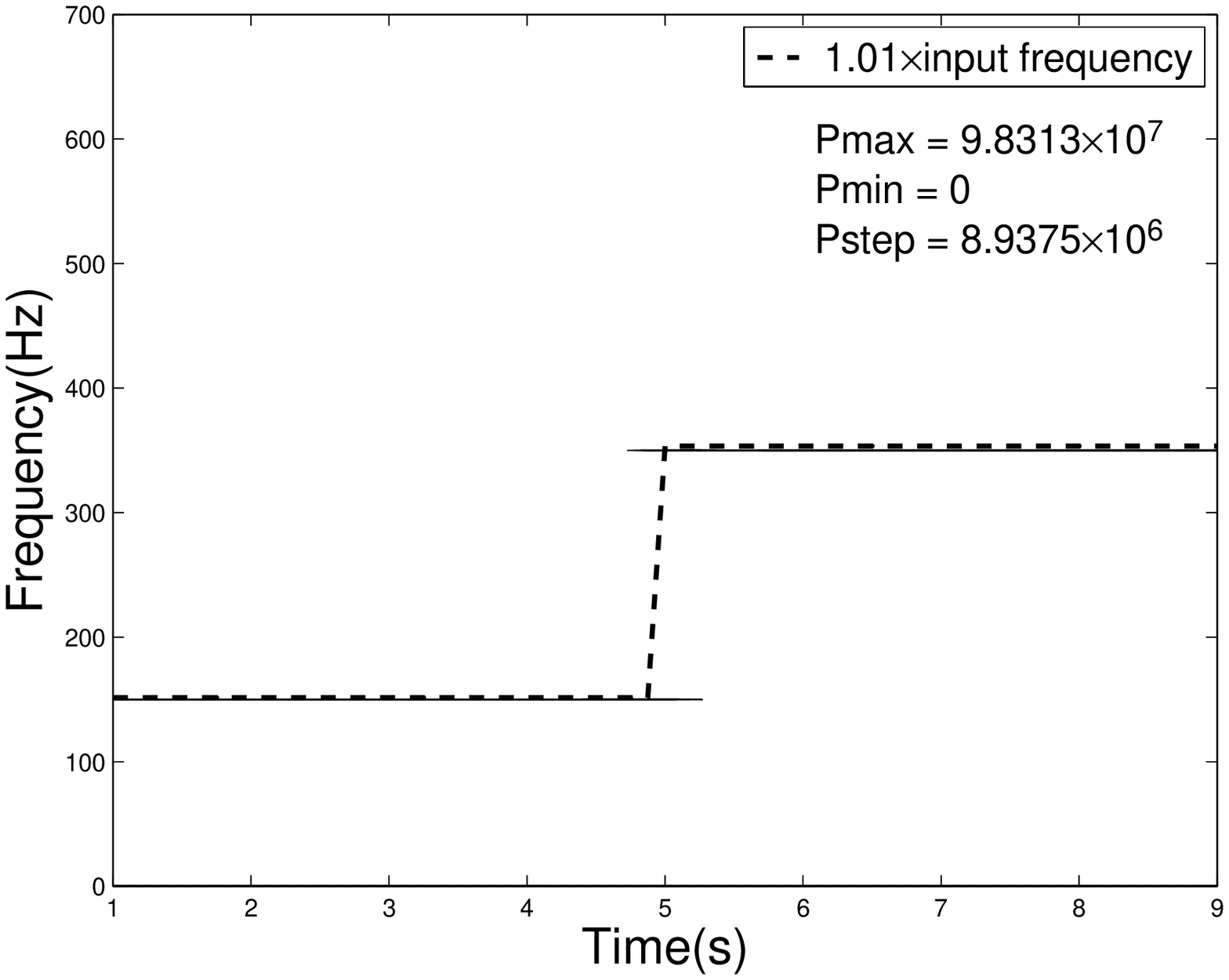}{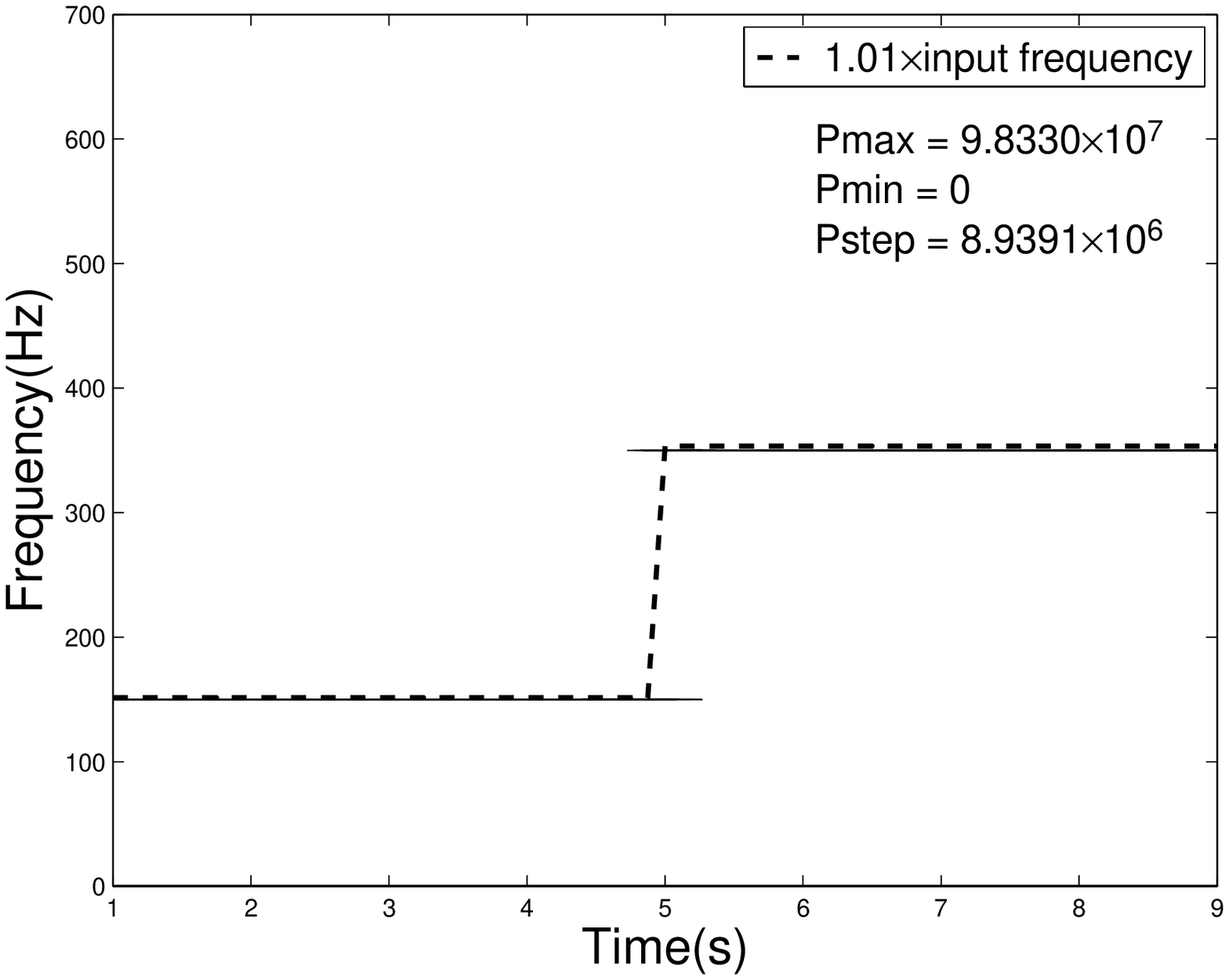} \caption{Calculated power spectra with
WFFT. The dashed line in each panel is for $1.01\times$ the
frequency of the input periodic signal. Left: the result of the
input periodic signal whose frequency jumps at 5 s as listed in
the third row of Table 1. Right: the result of the periodic signal
with frequency-jump and with Poissonian sampling as listed in the
forth row of Table 1.\label{fig-6}}
\end{figure}

\begin{figure}
\plottwo{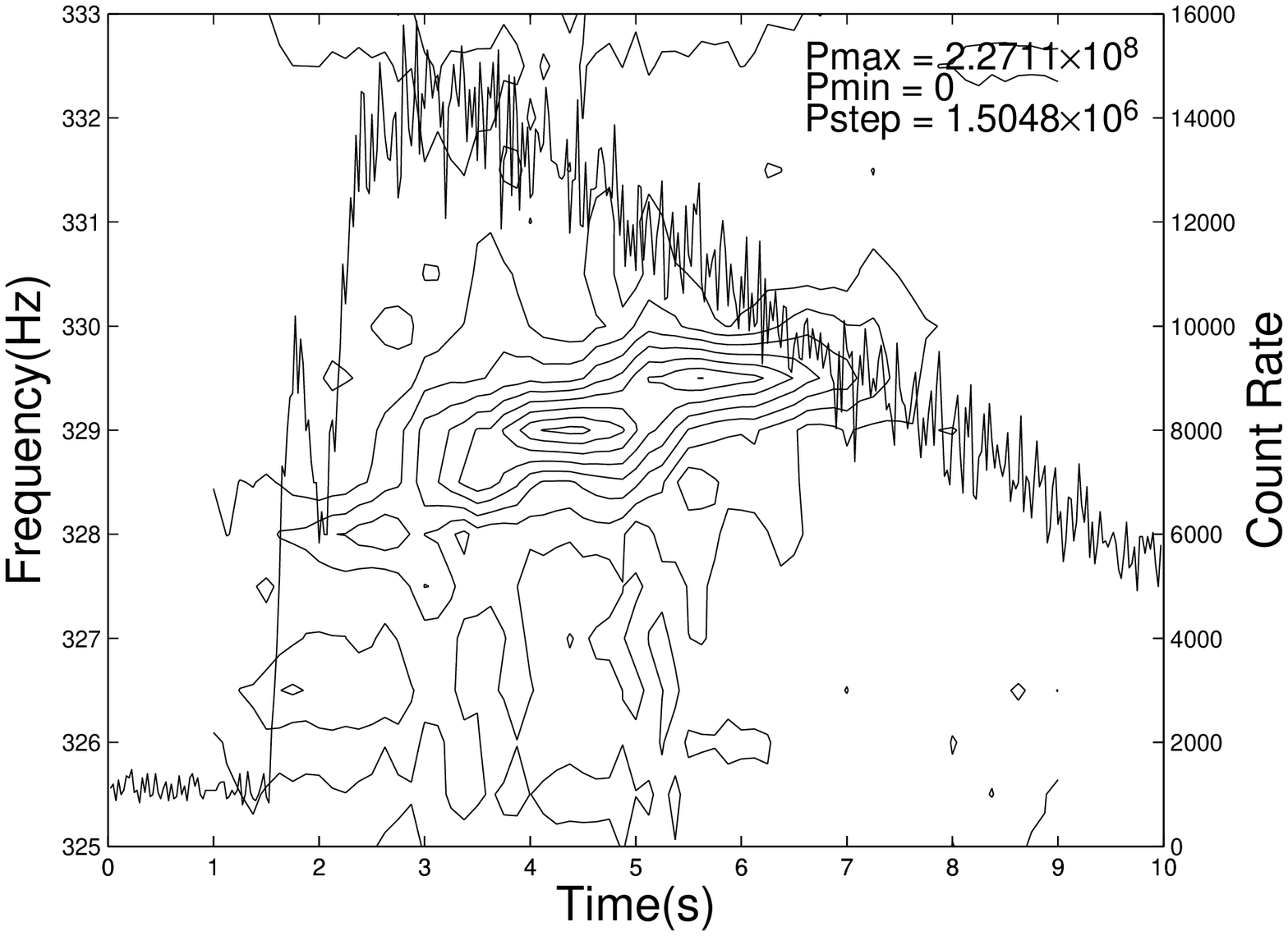}{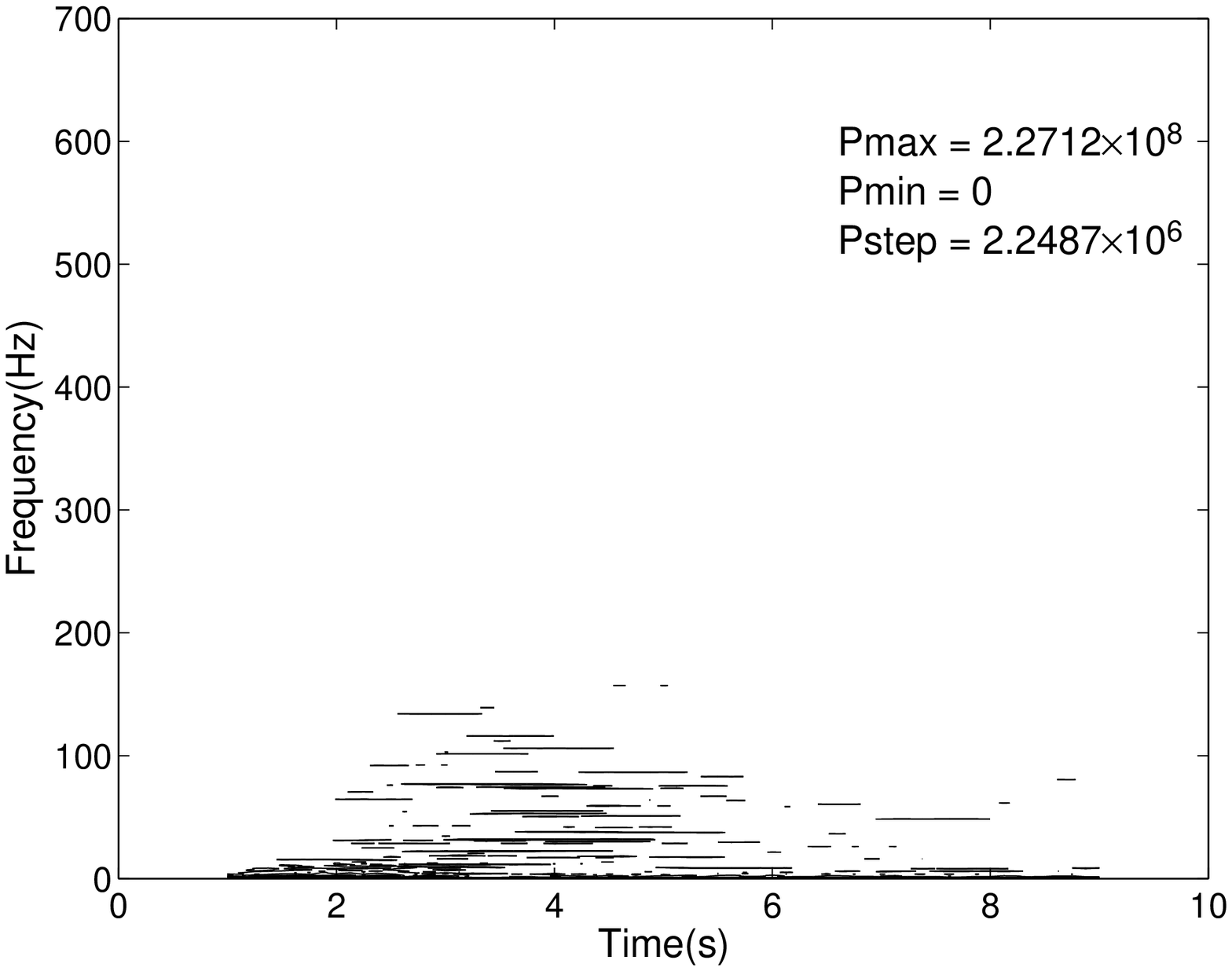} \caption{Left: light curve (solid line)
of the X-ray burst from 4U 1702-429 in 1997, and its power spectra
(contours) calculated with WFFT. The component with frequencies
about 328 Hz to 330 Hz has the most power. Right: power spectra
with WFFT for a smoothed light curve. The input time series is the
25-point smoothed light curve with Poissonian sampling as listed
in the first row of Table 2. The component with frequencies about
328 Hz to 330 Hz shown in the left panel are absent as a result of
smoothening.\label{fig-7}}
\end{figure}

\begin{figure}
\plottwo{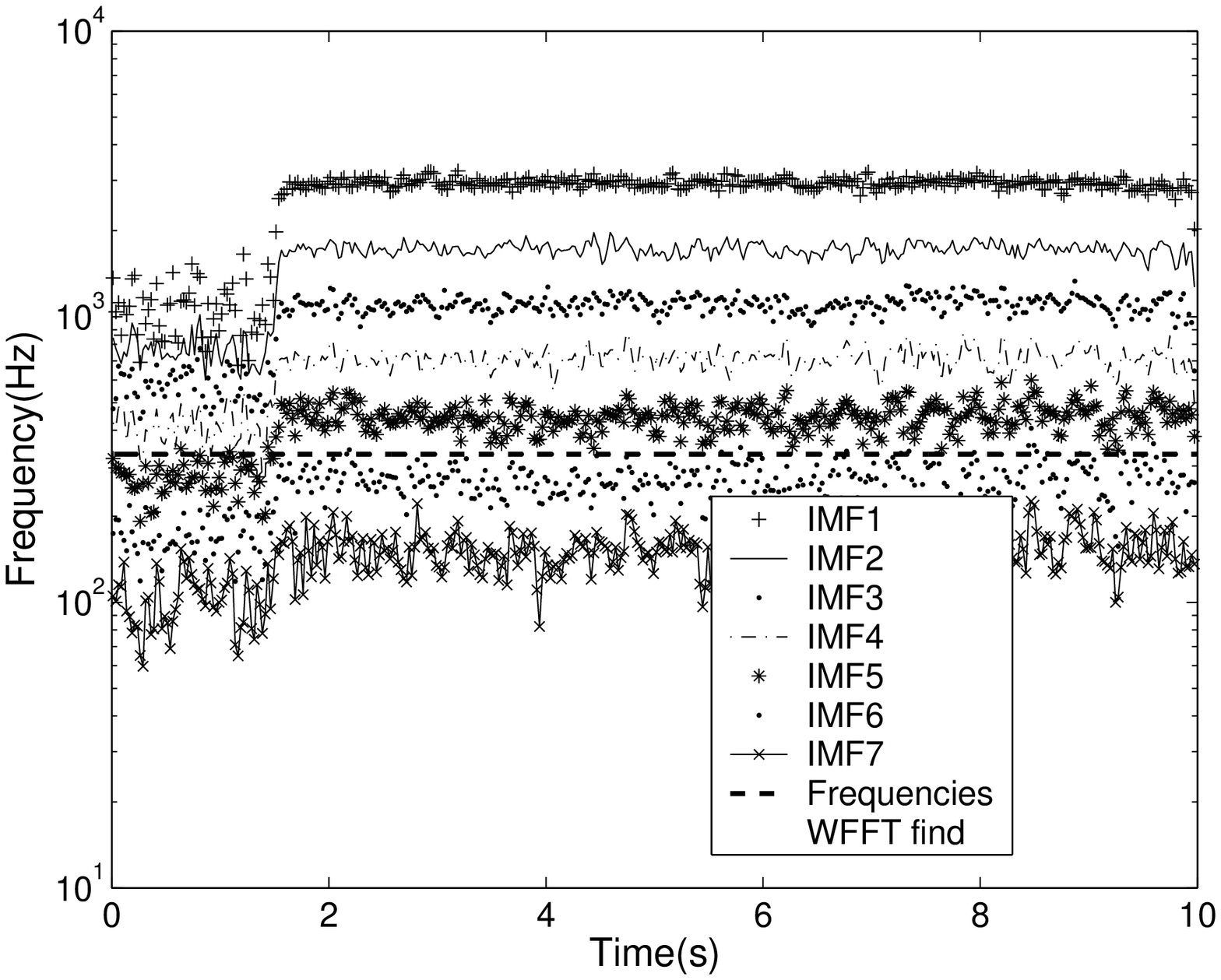}{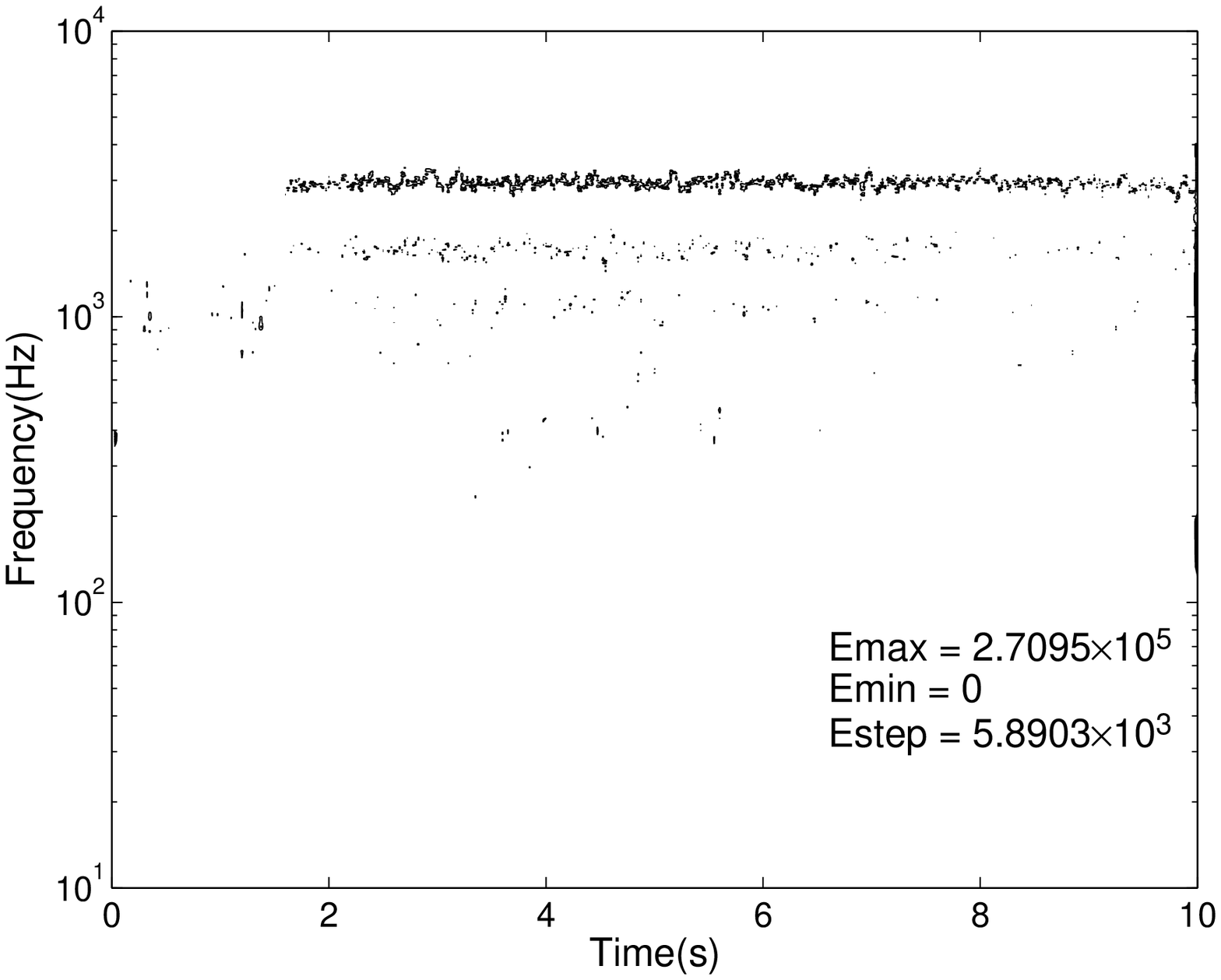} \caption{Calculated time-frequency
distribution (left) and energy-frequency-time distribution (right)
with HHT. The dashed line of the left panel is for the 330 Hz
frequency detected with the WFFT method. The input time series is
the light curve from 4U 1702-429 in 1997, the same as that of the
left panel of Figure 7. The HHT method cannot detect the frequency
component around 330 Hz.\label{fig-8}}
\end{figure}

\clearpage

\begin{figure}
\plottwo{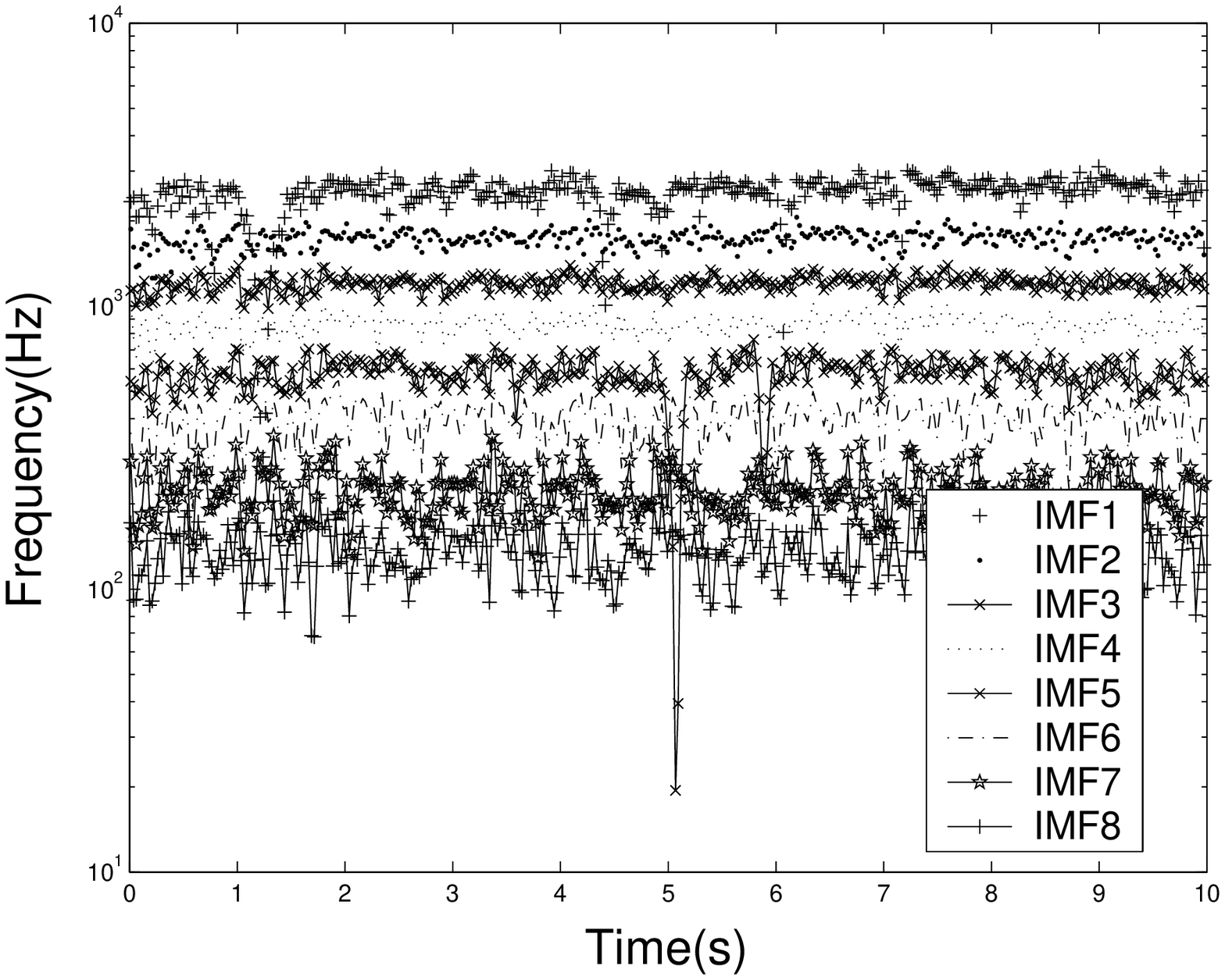}{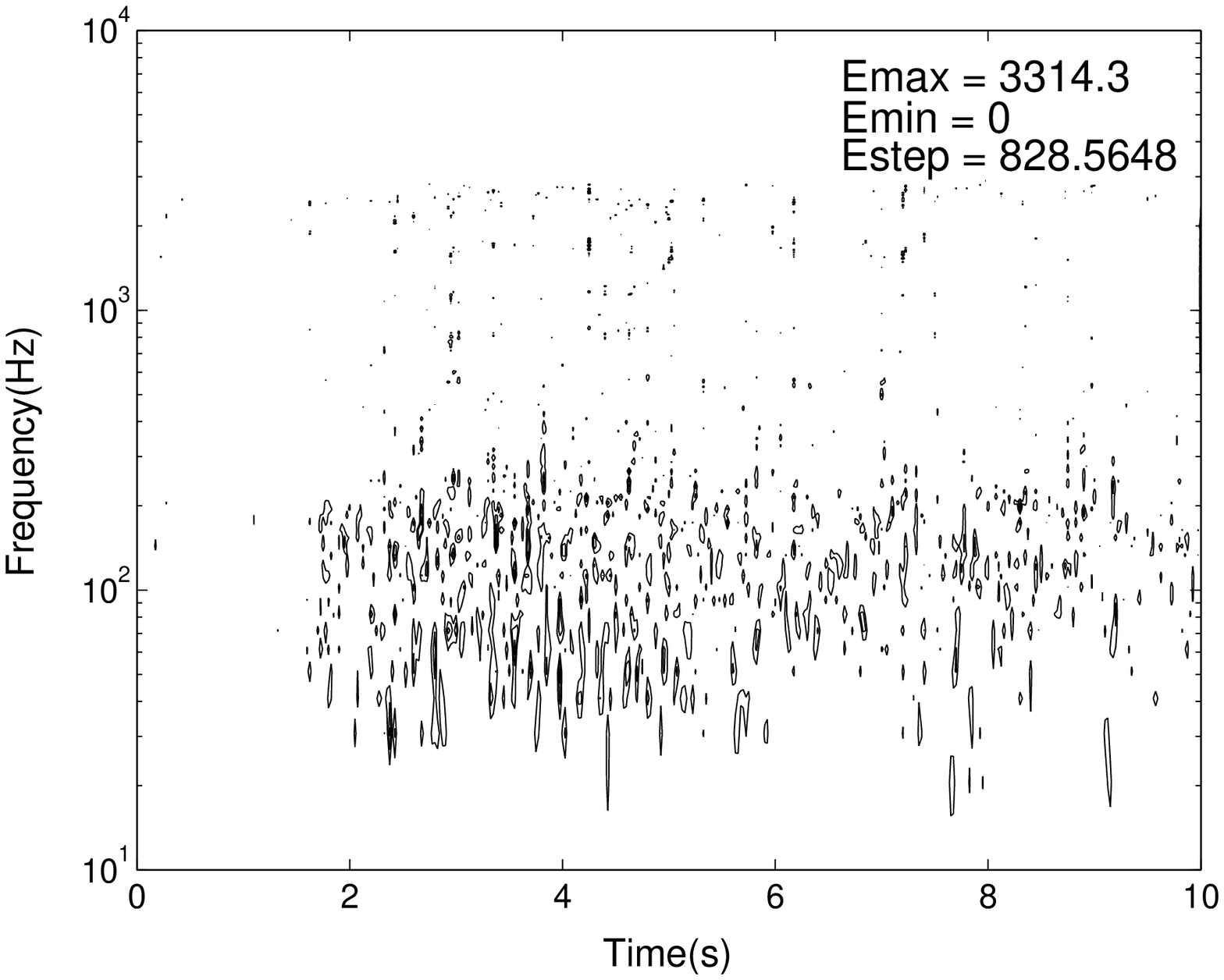} \caption{Calculated time-frequency
distribution (left) and energy-frequency-time distribution (right)
with HHT. The input time series is the 25-point smoothed light
curve with Poissonian sampling as listed in the first row of Table
2, the same as that of the right panel of Figure 7.\label{fig-9}}
\end{figure}

\begin{figure}
\plottwo{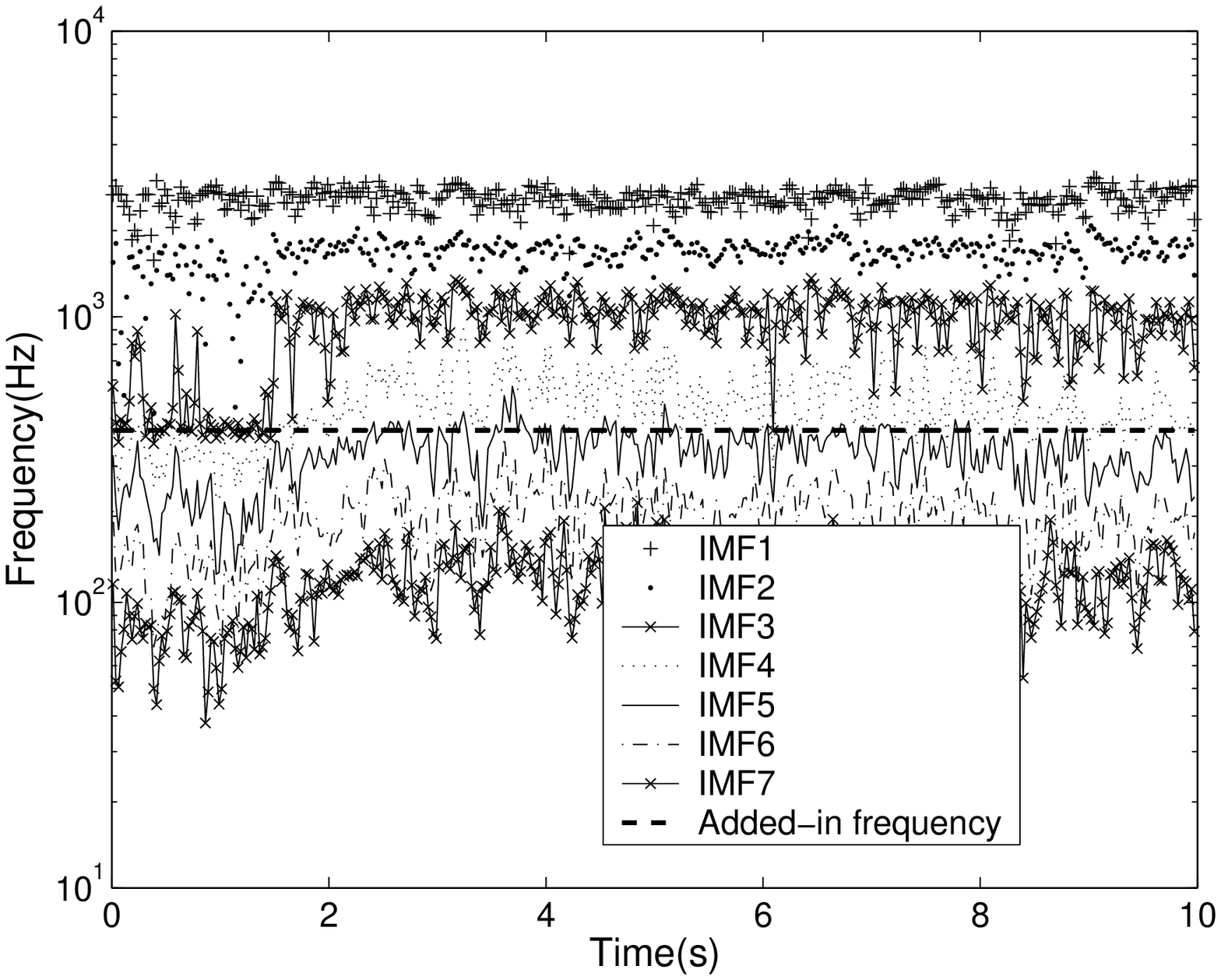}{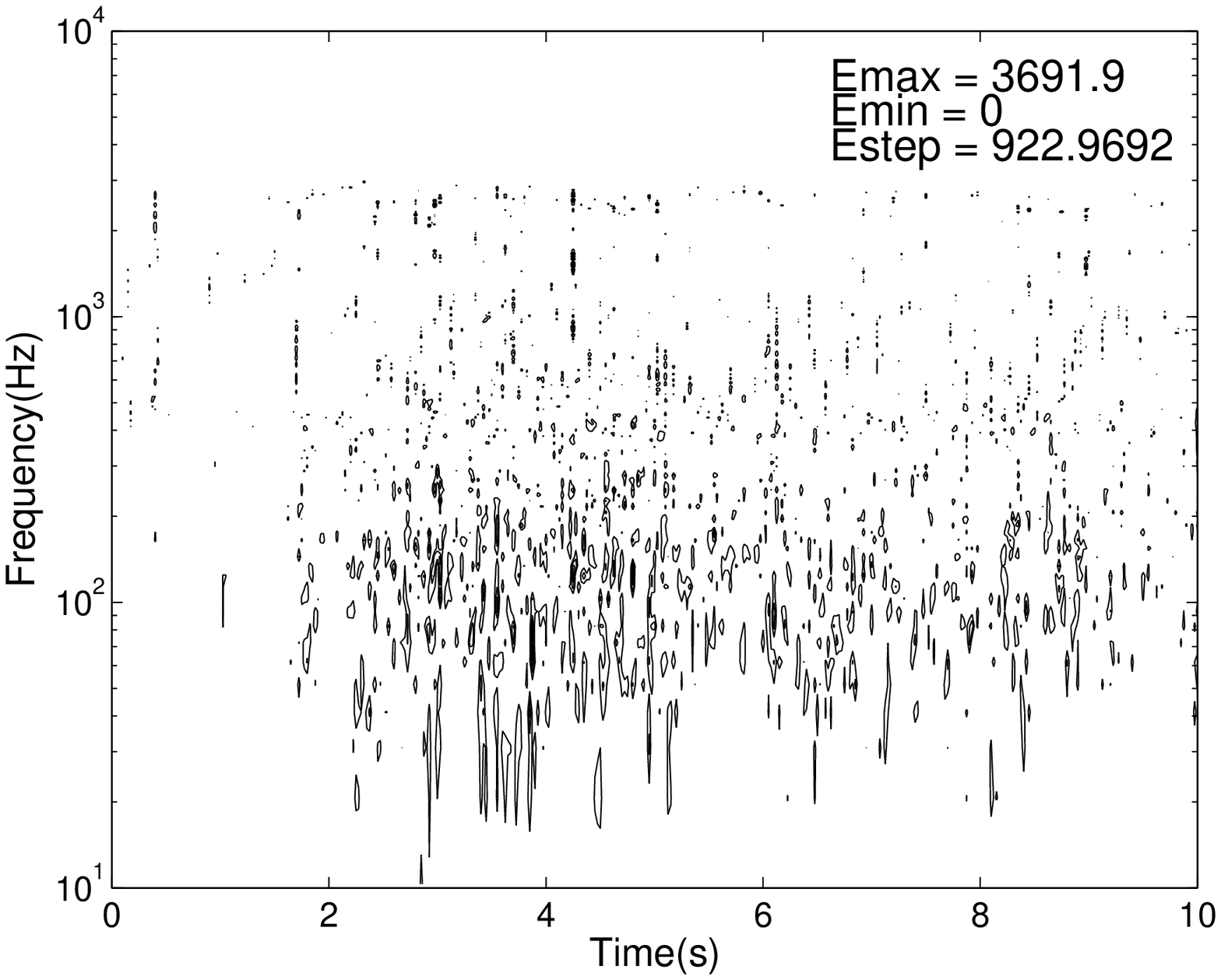} \caption{Calculated time-frequency
distribution (left) and energy-frequency-time distribution (right)
with HHT. The input time series is composed of the 25-point
smoothed light curve, added with a sinusoidal signal of 400 Hz and
the amplitude of 0.1 times the mean value of the light curve, and
then, Poissonian sampling as listed in the second row of Table 2.
The dashed line of the left panel is for the 400 Hz frequency of
the added-in sinusoidal signal.\label{fig-10}}
\end{figure}

\begin{figure}
\plottwo{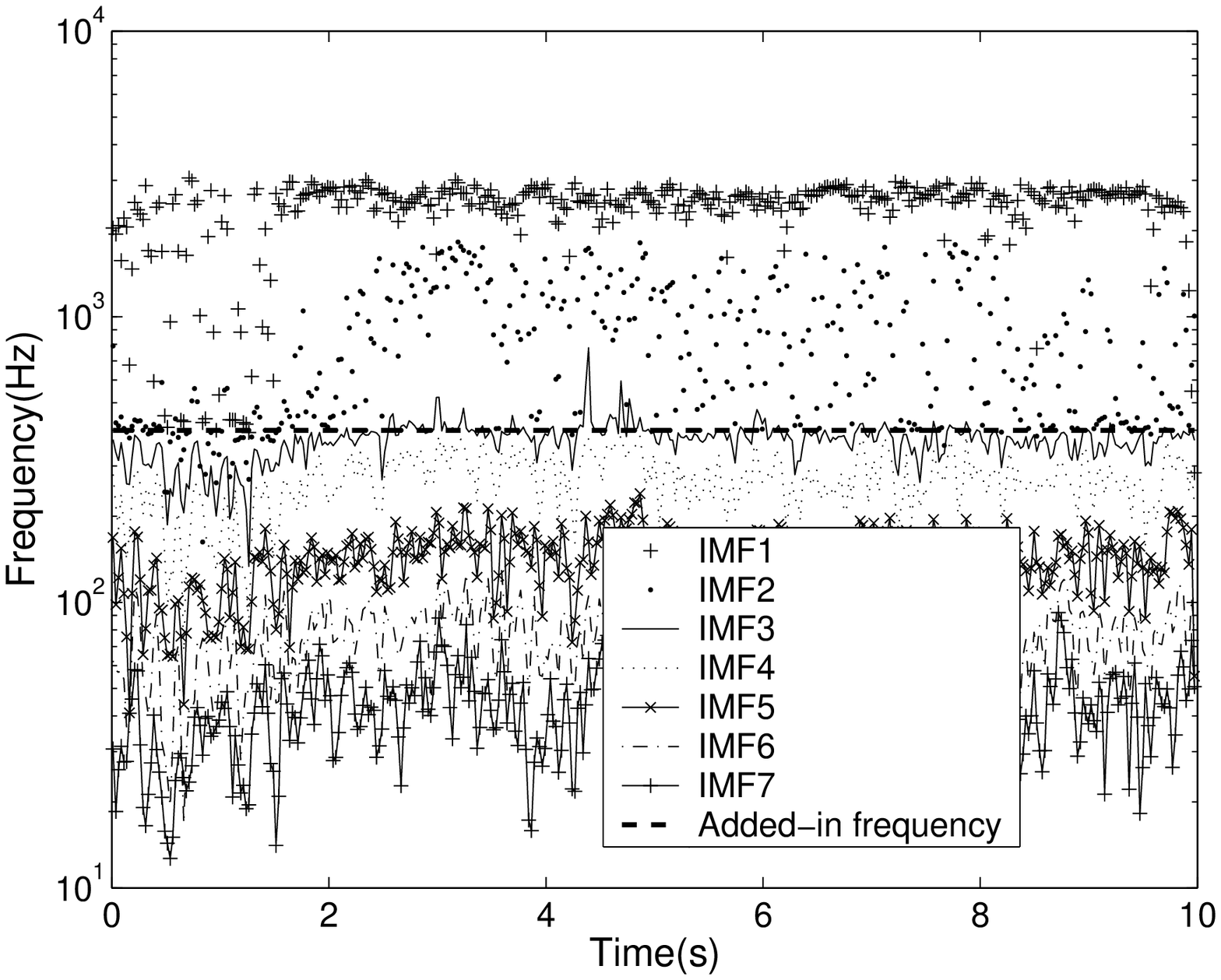}{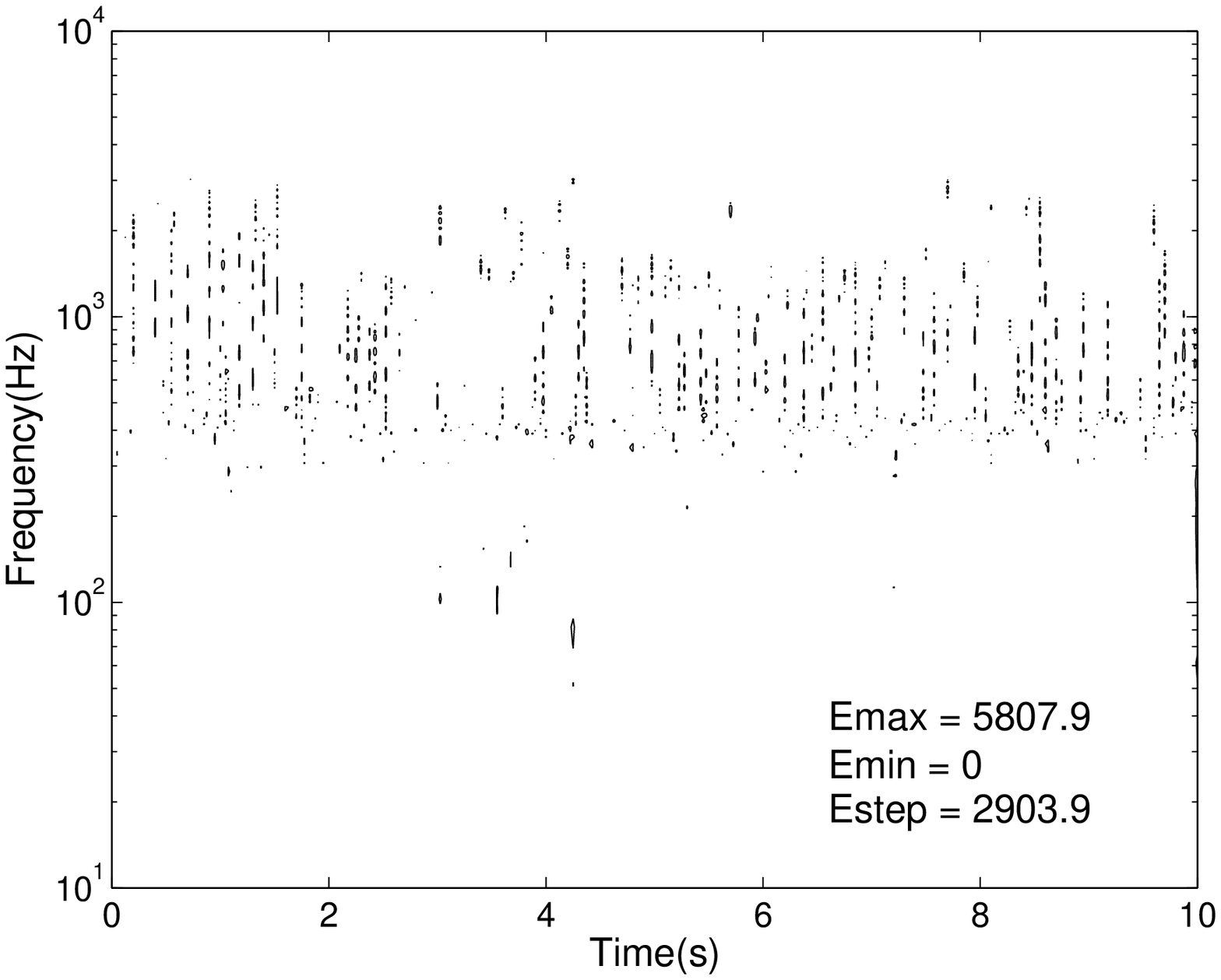} \caption{Calculated frequency-time
distribution (left) and energy-frequency-time distribution (right)
with HHT. The input time series is the same as that of Figure 10
but with a different amplitude of 0.4 times the mean value of
light curve, as listed in the third row of Table 2. The dashed
line of the left panel is for the 400 Hz frequency of the added-in
sinusoidal signal.\label{fig-11}}
\end{figure}

\begin{figure}
\plottwo{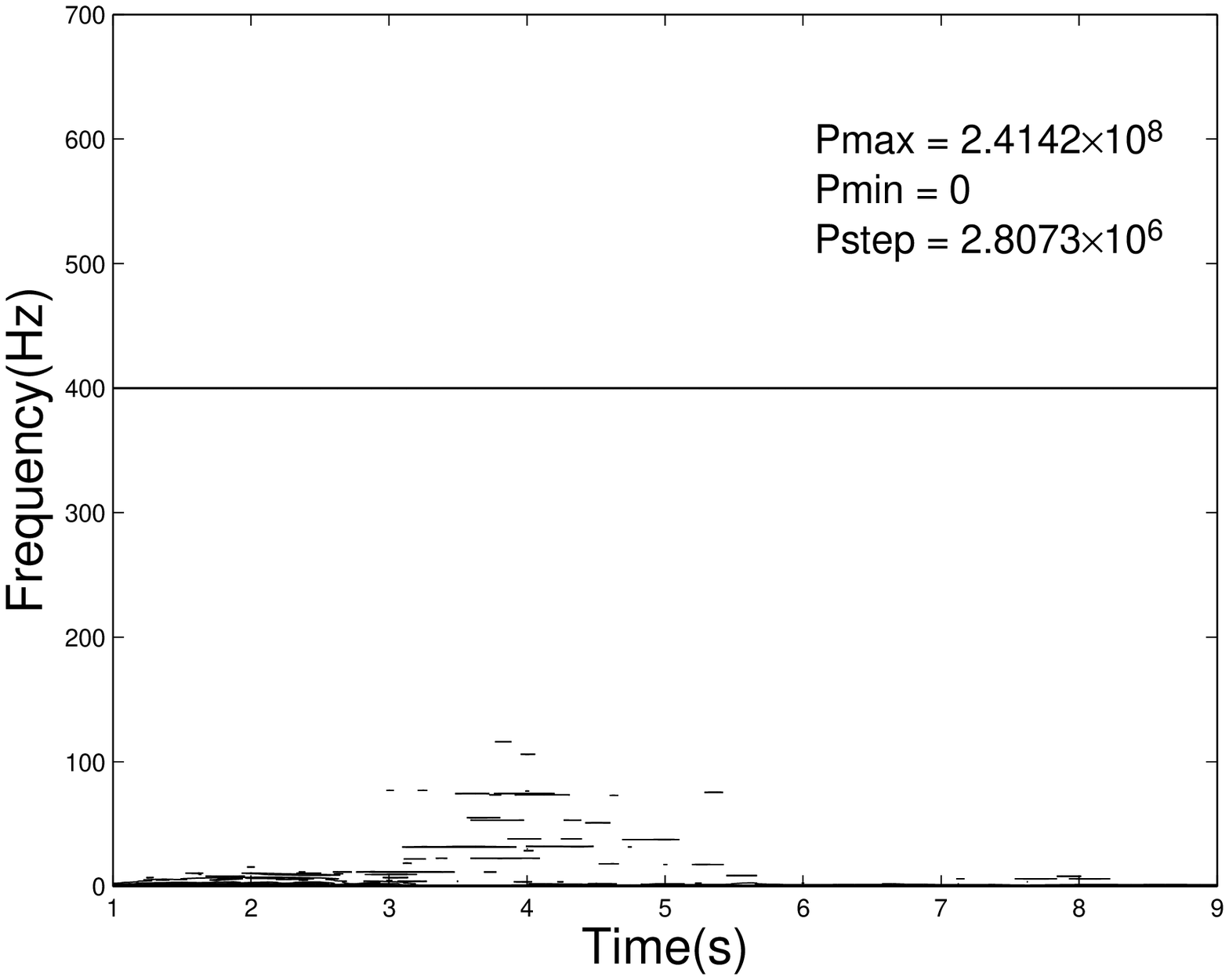}{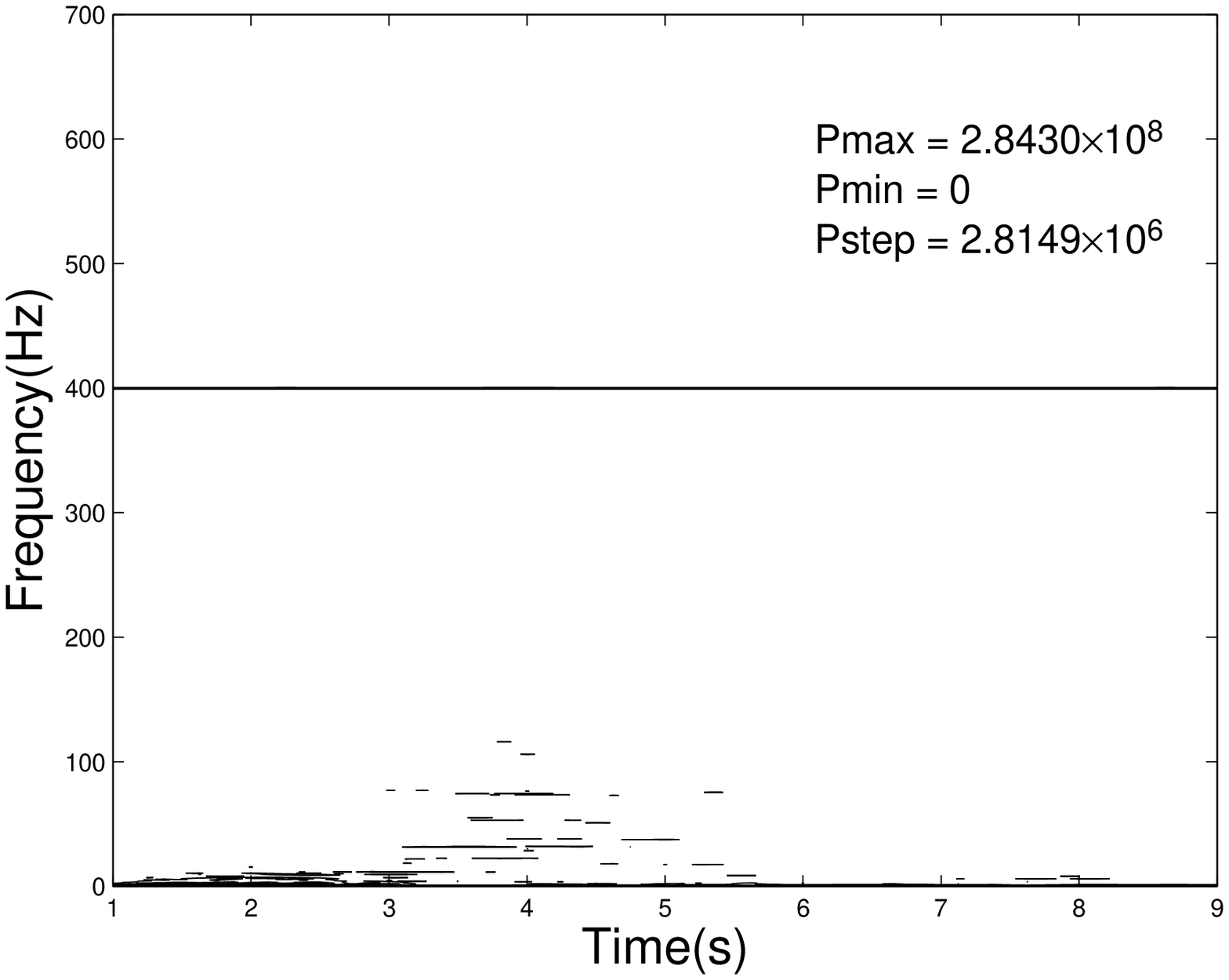} \caption{Power spectra with WFFT.
Left: the input time series is the same as that of Figure 10, as
listed in the second row of Table 2. Right: the input time series
is the same as that of Figure 11, as listed in the third row of
Table 2.\label{fig-12}}
\end{figure}

\end{document}